\documentclass{elsart}  %
\usepackage{epsfig}
\usepackage{amssymb}
\usepackage{cite}
\begin{document}
\begin{frontmatter}
\title{\hfill {\small Preprint HLRZ 84/96}\\ 
          Gauge-ball spectrum \\
        of the four-dimensional pure U(1) gauge theory}
\author{J. Cox, W. Franzki, J.~Jers{\'a}k}
\address{Institut f{\"u}r
Theoretische Physik E, RWTH Aachen, Germany}
\author{C.~B.~Lang}
\address{Institut f{\"u}r Theoretische Physik,
Karl-Franzens-Universit\"at Graz, Austria}
\author{T.~Neuhaus}
\address{Niels Bohr Institute, Univ. of Copenhagen, Denmark}
\author{P.~W.~Stephenson}
\address{DESY, Zeuthen, Germany}
\date{\today}
\begin{abstract}
  We investigate the continuum limit of the gauge-ball spectrum in the
  four-dimensional pure U(1) lattice gauge theory.  In the confinement
  phase we identify various states scaling with the correlation length
  exponent $\nu \simeq 0.35$. The square root of the string tension
  also scales with this exponent, which agrees with the non-Gaussian
  fixed point exponent recently found in the finite size studies of
  this theory.  Possible scenarios for constructing a non-Gaussian
  continuum theory with the observed gauge-ball spectrum are
  discussed. The $0^{++}$ state, however, scales with a Gaussian value
  $\nu \simeq 0.5$. This suggests the existence of a second, Gaussian
  continuum limit in the confinement phase and also the presence of a
  light or possibly massless scalar in the non-Gaussian continuum
  theory.  In the Coulomb phase we find evidence for a few
  gauge-balls, being resonances in multi-photon channels; they seem to
  approach the continuum limit with as yet unknown critical exponents.
  The maximal value of the renormalized coupling in this phase is
  determined and its universality confirmed.

\end{abstract}
\end{frontmatter}
\newpage

\section{Introduction}

Since studies of lattice gauge theories began it has been conjectured
that the pure U(1) gauge theory in four dimensions (4D) might possess
two continuum limits at the phase transition between the confinement
and the Coulomb phases \cite{Wi74Cr81,Pe78}, provided the transition
is of $2^{\rm nd}$ order. The continuum limit depends on the phase in
which it is approached.  In the confinement phase various gauge-balls
(GB) with finite masses, analogous to the glue-balls of pure QCD, as
well as a finite string tension $\sigma$, are expected.  In the
Coulomb phase with a massless photon dilute magnetic monopoles might
exist, whereas in the confinement phase they condense.  These
phenomena were clearly observed for a finite lattice cutoff $1/a$ long
ago \cite{DeTo80,BePa84}.  However, the continuum limit has remained
elusive for the subsequent 13 years.

The main reason for this delay has been the two-state signal at the
phase transition observed on finite lattices with periodic boundary
conditions \cite{JeNe83,EvJe85}. However, it has recently been
demonstrated that if, instead of such toroidal lattices, those with
trivial homotopy group are used, the two-state signal disappears for
the Wilson action \cite{LaNe94b}, extended Wilson action (defined
below) at $\gamma \le 0$ \cite{JeLa96a,JeLa96b}, and the Villain
action \cite{LaPe96}.  Furthermore, in high precision simulations on
specially constructed homogeneous spherical lattices, the finite size
scaling (FSS) behaviour turned out to be well described by the leading
term of the critical behaviour with a non-Gaussian value of the
correlation length critical exponent, $\nu = 0.365(8)$
\cite{JeLa96a,JeLa96b,LaPe96}.

Within the intrinsic uncertainties of numerical evidence this
demonstrates that the phase transition is of second order in some
parameter region and the corresponding continuum limit is governed by
a non-Gaussian fixed point. The expected continuum limit is thus
presumably a nontrivial quantum field theory which is not
asymptotically free.  Until now no such nonperturbatively defined
continuum quantum field theory has been established in four
dimensions. Further investigation of this theory is therefore of high
theoretical interest.

The use of spherical lattices has been crucial for the study of the
model by FSS methods, because this method combines scaling and finite
size phenomena. Thus the latter should not be distorted by topological
effects. However, once the critical properties are established, there
is no urgent need to use such complicated lattices for investigations
of the spectrum of the model.  The topology of finite lattices of
volume $V$ ought to be irrelevant for the thermodynamic limit, $V
\rightarrow \infty$, if that is taken before the continuum limit, when
the lattice constant $a \rightarrow 0$. The restrictions imposed by
the two-state signal on toroidal lattices can be avoided by using
large $V$ and choosing the coupling parameters outside the
metastability region on these lattices. In fact, fairly precise values
of $\nu = 0.28 - 0.42$ have been obtained by this approach in the
past.

In this work we therefore return to the 4D toroidal lattices and adapt
the latest methods of glue-ball measurements in QCD on such lattices
to the U(1) gauge group. Since the most recent studies have indicated
universality of the scaling behaviour \cite{JeLa96a,JeLa96b,LaPe96},
we have chosen an action reducing the two-state signal while keeping
the autocorrelation time reasonably short: the extended Wilson action
at the double-charge coupling $\gamma = -0.2$.

We have investigated in both phases on large lattices the correlation
functions in 20 channels with the zero momentum and in some channels
with the smallest nonzero momentum as well. We have also determined
the static potential from the same runs.

In the confinement phase, we have obtained masses $m_j$ (inverse
correlation lengths) in most channels. Within the considered parameter
range the finite size effects are negligible. Also the string tension
$\sigma$ has been determined from the values of the potential at the
largest separations.  We investigate the scaling behaviour of these
observables, i.e. their vanishing in lattice units, when $\beta$, the
standard Wilson coupling, approaches its critical value $\beta_c$.

Our main result for the confinement phase in the vicinity of the
critical point $\beta_c$ is the evidence for two groups of GB masses
with distinctly different scaling behaviour when the phase transition
is approached, $\tau = |\beta - \beta_c|\to 0$. Most of the GB
masses, and approximately also $\sqrt{\sigma}$, scale proportional to
$ \tau^{\nu_{\rm ng}}$, the value of the correlation length exponent
$\nu_{\rm ng}$ being non-Gaussian,
\begin{equation}
         \nu_{\rm ng} =  0.35(3).
\label{NUNG}
\end{equation}
This agrees with the exponent $\nu = 0.365(8)$ found in FSS studies on
spherical lattices \cite{JeLa96a,JeLa96b,LaPe96}.

However, the mass of the $J^{PC} = 0^{++}$ GB scales as 
$\tau^{\nu_{\rm g}}$, with
\begin{equation}
         \nu_{\rm g} =  0.49(7).
\label{NUG}
\end{equation}
This correlation length exponent agrees with the Gaussian value
$\nu = 1/2$. The critical point, common for both groups, is at
\begin{equation}
        \beta_c \simeq 1.1607(3).
\label{BETA_C}
\end{equation}
The errors indicated in the above three equations include both
statistical and systematic uncertainties of our results. Since the
latter are not independent, we also give the result for the ratio of
both exponents,
\begin{equation}
         \frac{\nu_{\rm ng}}{\nu_{\rm g}} =  0.71(8).
\label{NU_RATIO}
\end{equation}

These findings suggest that in the confinement phase two different
continuum limits are possible. One is non-Gaussian, in which all
states scaling as $\tau^{\nu_{\rm ng}}$ are obtained. Numerous GB
states with a somewhat degenerate spectrum, as well as the finite and
nonvanishing string tension should be expected. Our recent
\cite{JeLa96a,JeLa96b,LaPe96} and present results suggest that the
observables scaling as $\tau^{\nu_{\rm ng}}$ correspond to a
nontrivial continuum theory which is not asymptotically free. This
theory contains a light, possibly massless scalar.

The other continuum limit in the confinement phase, based on scaling
like $\tau^{\nu_{\rm g}}$, is presumably Gaussian. Of all the reliably
measured GB masses only that of the $0^{++}$ state has this behaviour.

The Gaussian exponent (\ref{NUG}) has not been detected in the finite
size scaling approach \cite{JeLa96a,JeLa96b,LaPe96}. However, the
existence of two eigenvalues of the linearized renormalization group
transformation matrix, consistent with the values (\ref{NUNG}) and
(\ref{NUG}) of the critical exponents, was observed back in the
eighties in Monte Carlo renormalization group studies of the model
\cite{Bu86,La86La87bLaRe87}. These exponents can now be interpreted
in terms of the scaling of physical observables.

In the Coulomb phase we find evidence that in the vicinity of the
phase transition, apart from the massless vector (``photon'') state,
also massive states are present. They appear to scale at the phase
transition. We identify them as resonances in two-photon and
three-photon channels, respectively, with significant finite size
effects.  As these are not yet fully under theoretical control, the
critical exponents could not be determined.  Vanishing of the
resonance masses when $\beta_c$ is approached would imply the
existence of a continuum limit with some dimensionful parameter in
this phase.

In the Coulomb phase we also obtain from the potential the
renormalized coupling $\alpha_{\rm R} = e^2_{\rm R}/4\pi$.  The value
of this coupling at the critical point $\beta_c$, $\alpha_{\rm R,c} =
0.19(1)$ is consistent with that in other formulations of the pure
compact U(1) gauge theory.  This is further support for the
conjectured universality of the renormalized coupling of this theory
in the Coulomb phase as one approaches the critical point
\cite{Ca80,Lu82,JeNe85}.

The results are presented as follows: In the next section we define
the model and describe its critical properties. In sec.~3 we discuss
the results for the static potential in both phases. Technical details
about the GB measurements are collected in sec.~4. In sec.~5 we then
present the results for the GB masses in the confinement phase.
Possible interpretations of their scaling behaviour are discussed in
sec.~6. Sec.~7 is devoted to the results of GB measurements in the
Coulomb phase. The conclusions and a long list of interesting open
questions are given in sec.~8. The appendix contains tables with more
detailed data.

\section{The pure U(1) lattice gauge theory}

\subsection{Action and phase diagram}

We consider pure compact U(1) lattice gauge theory with an extended
Wilson action,
\begin{equation}\label{ACTION}
         S = -\sum_P
              \left [\beta \cos(\Theta_P) + \gamma
                \cos(2\Theta_P)\right ].
\end{equation}
Here, $\Theta_P \in [0,2\pi)$ is the plaquette angle, i.e. the
argument of the product of U(1) link variables around a plaquette $P$,
and $\beta$ and $\gamma$ are the single and the double charge
representation couplings, respectively. Taking $\Theta_P =
a^2gF_{\mu\nu}$, where $a$ is the lattice spacing, and $\beta +
4\gamma = 1/g^2$, one obtains for weak coupling $g$ the usual
continuum action $S =\frac{1}{4} \int d^4xF_{\mu\nu}^2$.

This lattice gauge theory has a line of phase transitions between the
strong coupling confinement phase and the weak coupling Coulomb phase.
Its position for the Wilson action ($\gamma = 0$) is $\beta_c \simeq
1.011$ \cite{EvJe85}, whereas for $\gamma = -0.2$ it is given in
(\ref{BETA_C}). For $\gamma \ge +0.2$ the transition is clearly of
$1^{\rm st}$ order, weakening with decreasing $\gamma$
\cite{Bh82,EvJe85}.

The recent studies on spherical lattices strongly suggest that the
order changes at $\gamma = \gamma_0 \simeq 0$, $\gamma_0$ probably
being slightly positive, and is of $2^{\rm nd}$ order for $\gamma \le
\gamma_0$ \cite{JeLa96a,JeLa96b}. With decreasing $\gamma$ the $2^{\rm
  nd}$ order transition further weakens in the sense that the specific
heat peak decreases for fixed lattice size and the autocorrelation
time increases \cite{JeLa96b}. The scaling behaviour of bulk
quantities is universal at least in the range $-0.5 \le \gamma \le 0$.

On toroidal lattices the disturbing two-state signal weakens with
decreasing $\gamma$, but is present at least until $\gamma = -0.5$
\cite{EvJe85}.  For even smaller $\gamma$ the large autocorrelation
time makes simulations prohibitively expensive. Thus on the toroidal
lattices the two-state signal on the critical line cannot be avoided.

\subsection{Some earlier studies}

It has been suggested \cite{EvJe85} that the point $ \gamma_0$ is a
tricritical point (TCP). However, we would like to point out that the
TCP's are {\em not} defined merely as points on phase transition lines
where the change between the $1^{\rm st}$ and $2^{\rm nd}$ order takes
place. More is required: in a parameter space enlarged by an external
coupling to the order parameter three critical lines should emerge
from the TCP \cite{Gr73LaSa84,EvJe85}. This results in TCP's being
associated with tricritical exponents different from the critical ones
on the emerging critical lines. Up to now there has been no evidence
that the point $\gamma_0$ in the pure compact U(1) gauge theory is
really a TCP, because the other critical lines are not known.  The
problem is in finding the suitably enlarged coupling space while only
nonlocal order parameters distinguishing between the confinement and
Coulomb phases, like the string tension or the photon mass, exist.
This problem may be specific for gauge theories. It makes an
interpretation of the two observed scaling laws in terms of critical
and tricritical scaling behaviour uncertain.

Various investigations of the scaling behaviour of the model by
analytic means \cite{Ha81IrHa84}, the finite size scaling analysis
\cite{Bh82,LaNa80Bh81MuSc82} and the Monte Carlo renormalization
group (MCRG) method \cite{Bu86,La86La87bLaRe87,GuNo86Ha88} found
the correlation length exponent $\nu$ to be in the range $\nu \simeq
0.28 -0.42$.  In some MCRG studies the existence of a second value of
this exponent, consistent with $\nu = 0.5$, was observed
\cite{Bu86,La86La87bLaRe87} suggesting that one of these exponents
is tricritical and the other critical. In ref.~\cite{Bu86} the
tricritical value was suggested to be $\nu = 0.5$, whereas in
refs.~\cite{La86La87bLaRe87} the assignment was opposite, the
tricritical exponent being the non-Gaussian one.  Recently, evidence
has been provided, that in the coupling space enlarged by monopole
coupling, a second order phase transition with non-Gaussian critical
exponents may be observed \cite{KeRe97b}.

Because of the doubts as to whether a continuum limit can be obtained,
calculations of the spectrum of the model are sporadic. The massless
photon in the Coulomb phase was observed in various studies
\cite{BePa84,EvJa87b,CoHe87,NaPl91,BoMi93b} and the presence of
several gauge-balls in both phases was indicated in \cite{BePa84}. In
\cite{CoHe87,NaPl91} the $1^{--}$ gauge-ball, the ``massive photon''
in the confinement phase was found and in \cite{CoHe87} its scaling
investigated and found to be consistent with $\nu = 0.33$. The
monopole mass was measured in the Coulomb phase in a dual formulation
of the theory \cite{PoWi91}. To our knowledge its scaling behaviour
has not yet been studied.

\subsection{Overview of our measurements}

Motivated by the recent progress in understanding the continuum limit,
in the present work we investigate the spectrum and its scaling
behaviour. Our choice of $\gamma = -0.2$ is a compromise between the
requirements of minimizing both the two-state signal and the
autocorrelation time.

In the simulations we have used a vectorized three-hit Metropolis
algorithm with acceptance around 50\% at every hit. The measurements
have been performed every 25 sweeps. The used lattice sizes $L_s^3
L_t$, $L_t = 2 L_s$, are listed in table \ref{DATA_POINTS}. In the
confinement phase the presented data have been obtained on the $L_s^3
L_t = 16^3 32$ and $20^3 40$ lattices (smaller lattices being used
only for explorative calculations).
In the Coulomb phase the lattice size has been varied in a broad range.

\begin{table}[tbp]
  \hfuzz=1pt
  \begin{center}
    \leavevmode
    \footnotesize
    \begin{tabular}{|c|c|c|c|c|c|c|c|}\hline
      $\beta$&$8^316$&$10^320$&$12^324$&$14^328$&$16^332$&$18^336$&$20^340$\\
      \hline  
      $1.100$ & -& - & 6.0&-  & 5.2& -  & - \\
      $1.130$ & 8.0& - & 6.0&-  & 5.7& -  & - \\
      $1.135$ & -& - & -&-  & 4.5& -  & - \\
      $1.140$ &5.0 & - & 6.0&-  & 6.8& -  & - \\
      $1.145$ & -& - & -&-  & 6.6& -  & - \\
      $1.150$ &16.0& - & 9.0&-  & 7.5& -  & - \\
      $1.152$ & 5.0& - & 5.0&-  & 3.6 & -  & - \\
      $1.154$ & -& - & 9.0&-  & 6.4& -  & 2.4\\
      $1.156$ & -& - & 2.0&-  & 6.4& -  & 1.6\\
      $1.158$ & -& - & -&-  & 6.1& -  & 2.0\\
      $1.159$ & -& - & -&-  &  1.0& -  &13.7\\
      $1.160$ & -& - & -&-  & -  & -  & 0.6\\
      \hline
      $1.161$ &10.0& 4.0&8.0& 1.0& 6.0 & -  & 1.6\\
      $1.162$ &5.0 & 4.0& 4.0& 1.0& 3.0 & 1.6 & 1.2\\
      $1.165$ &52.0& 4.0&10.0& 1.0& 14.0& 0.8 & 2.4\\
      $1.170$ &44.0&22.0&14.0&9.4& 16.7& 2.5 & 3.0\\
      $1.180$ &14.0& 4.0& 4.0& 1.0& 10.0& 0.8 & 1.6\\
      $1.190$ & -& -& - & - & 2.0 & -  & - \\
      $1.200$ &5.0 & -&4.0 & - & 5.0& -  & 1.0\\
      \hline  
    \end{tabular}
  \end{center}
  \caption
  {List of our data in the confinement (upper part) and Coulomb
    phases. The numbers in the table are the
    numbers of measurements in multiples of 1000.}
  \label{DATA_POINTS}
\end{table}


In table \ref{DATA_POINTS} we list the $\beta$ values at which the
measurements were made. The point $\beta = 1.100$ turned out to be too
deep in the confinement phase to allow reliable measurements of the
masses. At $\beta = 1.160$ an indication of phase flips was observed
even on our largest lattice. These two points have therefore been
excluded from the data analysis, leaving 10 points $\beta = 1.130 -
1.159$ in the confinement phase. In the Coulomb phase the data have
been collected at 7 points. All data in the vicinity of $\beta_c$ have
been checked for absence of any indication of phase flips in the time
evolution of the plaquette energy.

The calculation of the GB masses at many $\beta$ points, required for
the study of their scaling behaviour, limited the statistics at any
given $\beta$. This made it difficult to distinguish GB states from
background in channels with weak GB signal. To improve this situation
we have accumulated substantially higher statistics for at least one
point, $\beta = 1.159$ on the $20^3 40$ lattice. This is the point
closest to $\beta_c$ that still has no phase flips.  This allows us to
determine the GB masses in channels with weaker signal with higher
reliability at least at one $\beta$.

As listed in table \ref{DATA_POINTS}, we have typically accumulated
several thousand statistically independent gauge field configurations.
We have made two sorts of measurements. The expectation values of the
rectangular Wilson loop operators $W_{R,T}$ with temporal extension
$T$ and extension $R$ in any of the spatial directions have been
determined.  All Wilson loops of sizes $R=1,...,L_s/2$ and
$T=1,...,L_t/2$ have been considered. From this we have obtained the
static potential essentially by standard methods.

Our main task, the measurement of the GB masses in various channels at
various $\beta$, has been performed by adopting the latest methods of
glue-ball measurements in pure gauge lattice QCD to the U(1) gauge
group. We have followed \cite{Te86Te87,MiTe89} in implementing the
techniques of smearing and diagonalization of the correlation
matrices. A more detailed description is given in sec.~4. The
effective energies $\epsilon_j(t)$ in various GB channels $j$ with
zero and, in some cases, smallest nonzero lattice momentum have been
obtained at various distances $t$. The GB masses $m_j$ have been
obtained from the plateaus of these energies where possible.  In some
channels with weak GB signal $\epsilon_j(t)$ at only one $t$, $t = 1$
could be used, giving at least an estimate of the GB mass. In this
procedure the standard lattice dispersion relation has been used.

The lattice calculations provide static potential $V(R)$, string
tension $\sigma$, and masses $m_j$ in the lattice units. To give these
observables in physical units, e.g.  $m_j^{\rm phys} = m_j/a$, it is
necessary to specify the value and the vanishing of the lattice
constant $a$. As several scenarios for the continuum limit $a
\rightarrow 0$ must be considered, we postpone this issue to sec.~6
and until then use the lattice units only.

Simultaneously with the above measurements we have also determined
various fermionic observables in the quenched approximation. This is
aimed at a study of the U(1) theory with staggered fermions and will
be published separately \cite{CoFr97c}.

\section{Static potential}

\subsection{Measurement of the static potential}

In the static potential analysis we include Wilson loop expectation
values whose noise to signal ratio is smaller than 20\% of the
expectation value itself.  Typically we observe that Wilson loop
expectation values of magnitude $0.0001$ and larger are measurable
within this threshold, entirely independent of $\beta$.

The potential $V(R)$ between static charges is defined as
\begin{equation}
                 V(R)=\lim_{T\to\infty}{{1}\over{T}}
\ln\langle W_{R,T}\rangle .
\end{equation}
For finite systems it must be specified in a suitable way.

In the confinement phase the Wilson loop expectation values decay
rapidly with increasing $T$. To extend the usable $T$ interval we use
the Ansatz
\begin{equation}
     \langle W_{R,T}\rangle
          =C(R) e^{-V(R)T}+C_1(R) e^{-V_1(R)T}~,~~~V(R) < V_1(R),
\label{wilson_loop_fit_form}
\end{equation}
at fixed $R$. This allows for two leading eigenvalues in the transfer
matrix representation of the gauge invariant correlation functions.
When compared to the data this truncation appears to be a reasonable
approximation. The four-parameter fit at each R is performed solely
for those values of $R$ where at least 6 data points are below the 20
percent threshold in a $T$-interval starting with $T=1$.  We find that
$V_1(R)$ exceeds the static potential $V(R)$ by a factor of about $2$
throughout the confinement phase. At $\beta=1.130$, the data point
deepest in the confinement phase, we are only able with this method to
determine the static potential at distances $R=1,2,3,4$. The situation
gradually improves as the critical point is approached. At
$\beta=1.159$, the data point closest to criticality, we were able to
determine the static potential at distances $1\leq R\leq 7$.  We have
checked that the values of $V(R)$ obtained in this way are consistent
with those obtained by using the $T\ge R$ loops only and setting
$C_1(R) = 0$.

In the Coulomb phase the Wilson loops of the considered sizes are much
better measurable. There a fit to the form
(\ref{wilson_loop_fit_form}) with $C_1(R) = 0$ for values of $T \ge 8$
unambiguously determines the static potential for $1\leq R\leq 8$ on
$16^3 32$ lattice.

\subsection{Confinement phase}

In the confinement phase we are primarily interested in the string
tension $\sigma$ and its scaling behaviour. The string tension is
defined through the asymptotic behaviour of the potential $V(R)\propto
\sigma R$ as $R \rightarrow \infty$.  In order to extract $\sigma$ at
the available distances $R$, a reliable parameterization of the short
distance part of the potential would be of much help, similar to
lattice QCD. In the U(1) theory we face a serious obstacle: the lack
of theoretical knowledge of the potential at short distances in the
strongly interacting theory which is not asymptotically free.

We have experimented with various fits to the potential on the whole
available $R$-interval, parameterizing the short distance behaviour of
the potential in several ways. The lattice Coulomb potential of a
massless exchange particle, as well as the lattice Yukawa potential of
a massive exchange particle were considered. At the available
distances these Ans\"atze influence the values of $\sigma$
significantly. This effect is at the 10 percent level for our smallest
$\beta$-values, where the potential is only known at few $R$-values,
and is less pronounced, at the level of a few percent, close to the
critical point. This is so also if the gradient of the
potential, the force, is considered. In summary, we cannot avoid a
dependence of the fitted string tension values on the Ansatz for the
short range part of the potential if all the data for $V(R)$ are
fitted. As this systematic error decreases when the critical point is
approached, it distorts the scaling behaviour of $\sigma$ determined
in such a way.

In this situation we expect that a straight line fit to the potential
for the largest $R$-values results in a less ambiguous and model
independent determination of the string tension. For the final
determination of $\sigma$ at all the $\beta$ values considered we used
the potential data at the 3 largest $R$ values with $V(R)$ being
measurable in the sense described above. In table \ref{TAB_SIGMA} the
obtained values of the string tension in the confinement phase are
listed.

\begin{table}[tbp]
  \hfuzz=1pt
  \begin{center}
    \leavevmode
    \footnotesize
\begin{tabular}{|c|ll|}\hline
  $\beta$  &  \multicolumn{2}{c|}{$\sigma$}  \\
           & \multicolumn{1}{c}{$16^3 32$} & \multicolumn{1}{c|}{$20^3
           40$} \\ 
               \hline
   1.130   & 0.268(6)  &            \\
   1.135   & 0.244(5)  &            \\
   1.140   & 0.226(2)  &            \\
   1.145   & 0.178(4)  &            \\
   1.150   & 0.137(6)  &            \\
   1.152   & 0.109(10) &            \\
   1.154   & 0.118(11) &  0.139(15) \\
   1.156   & 0.086(14) &  0.082(13) \\
   1.158   & 0.066(9)  &  0.062(17) \\
   1.159   &           &  0.028(13) \\ 
             \hline
\end{tabular}
\end{center}
\caption{
Values of the string tension $\sigma$ in the confinement phase.
}
\label{TAB_SIGMA}
\end{table}


The square root of the string tension $\sqrt\sigma$ is expected to
scale like a mass. We fit the string tension data using the critical
value $\beta_c=1.1607$ which is obtained below from the scaling of the
GB spectrum. Recalling that our string tension determination is less
affected by systematic uncertainties close to the phase transition we
restrict the fit to $1.14 \le \beta < \beta_c$. The fit to the scaling
form
\begin{equation}
     \sqrt{\sigma}=c_{\sigma}\tau^{\nu_\sigma}
\label{SIGMA_SCALING}
\end{equation}
results in the exponent value $\nu_\sigma=0.35(2)$ and a
$\chi^2/N_{\rm{DF}}$-value for the fit of about $1.9$. We test the
stability of the fit by the omission of further data points and obtain
$\nu$-values scattering in a 10 percent interval around the quoted
value. The square root of the string tension data and the fit are
displayed in fig.~\ref{fig:alphasigma}a.  Within the present numerical
and systematic uncertainties we conclude that the string tensions
scaling behaviour is consistently described by the non-Gaussian
exponent $\nu_\sigma=\nu_{ng}$.

\subsection{Coulomb phase}

The static potential in the Coulomb phase is well known to be
dominated by the exchange of the massless photon. Its determination
and a confirmation of its Coulomb form is a spin-off of our study
of the spectrum in this phase.  Nevertheless, it is of interest to
determine the maximal renormalized coupling at $\gamma = -0.2$, as
this has not been done before.

An appropriate representation of the lattice Coulomb potential
$V_{\rm L,C}$ is obtained by the lattice Fourier transform of a massless
bosonic propagator. It has the form
\begin{equation}
  V_{\rm L,C}(R)=\frac{4\pi}{L_{\rm s}^3}\sum_{\vec{k}\not= 0}
\frac{{\rm e}^{{\rm i} k_1R }}
  {\sum_{j=1}^3 2(1-\cos(k_j))},\quad k_j = 0, \frac{2\pi}{L_{\rm s}},
  \dots,\frac{2\pi(L_{\rm s}-1)}{L_{\rm s}}.
\end{equation}
We have fitted the static potential at seven $\beta$-points in the
Coulomb phase with the form
\begin{equation}
        V(R)=-\alpha_{\rm R} V_{\rm L,C}(R)+ \sigma R + C
\label{coulomb_phase_potential}
\end{equation}
and obtain a good description of the potential. The coefficient
$\sigma$ is consistent with zero within error bars, confirming the
vanishing of the string tension in the Coulomb phase.  We therefore
omit the linear term altogether in the determination of the
renormalized fine structure constant $\alpha_{\rm R}$. The values
of $\alpha_{\rm R}$ grow when the critical point is approached. Our
data are collected in table \ref{TAB_ALPHA}.

\begin{table}[tbp]
  \hfuzz=1pt
  \begin{center}
    \leavevmode
    \footnotesize
\begin{tabular}{|c|cc|}\hline
 $\beta$  & $L_s^3L_t$ & $\alpha_R$ \\ \hline
  1.161   & $20^3 40$  &   0.180(5) \\
  1.162   & $20^3 40$  &   0.186(3) \\ 
  1.165   & $16^3 32$  &   0.179(4) \\
  1.170   & $16^3 32$  &   0.169(3) \\
  1.180   & $16^3 32$  &   0.162(2) \\
  1.190   & $16^3 32$  &   0.157(2) \\
  1.200   & $16^3 32$  &   0.152(2) \\ \hline
\end{tabular}
\end{center}
\caption{
  Values of the renormalized fine
  structure constant $\alpha_R$ in the Coulomb phase. }
\label{TAB_ALPHA}
\end{table}


Based on the analogy between the low temperature (small coupling)
expansions of the $D=2$ XY spin model in its spin-wave phase, and of
the compact U(1) gauge theory in the Coulomb phase, one expects
\cite{Lu82} that the renormalized fine structure constant behaves
according to
\begin{equation}
          \alpha_{\rm R}(\beta)=\alpha_{\rm R,c} 
            - A_{\alpha}(\beta-\beta_c)^{\lambda}~,
             ~\beta>\beta_c. \\
\label{alfa_scaling}
\end{equation}
At the critical point a finite value of $\alpha_{\rm R,c}$ has been
predicted and it has been suggested that this value might be universal
\cite{Ca80,Lu82}. This is in complete analogy to the $D=2$ XY model,
where the corresponding coupling of the $2$-dimensional analog of the
Coulomb potential has the critical value $1/4$.

Previous simulations \cite{JeNe85} of the model with the Wilson and
Villain actions in the Coulomb phase showed consistency between the
data and (\ref{alfa_scaling}): for both actions obtained the value
$\alpha_{\rm R,c} \simeq 0.20$ with an error about 0.02.

\begin{figure}[htb]
  \centerline{
    \hbox{
      \psfig{file=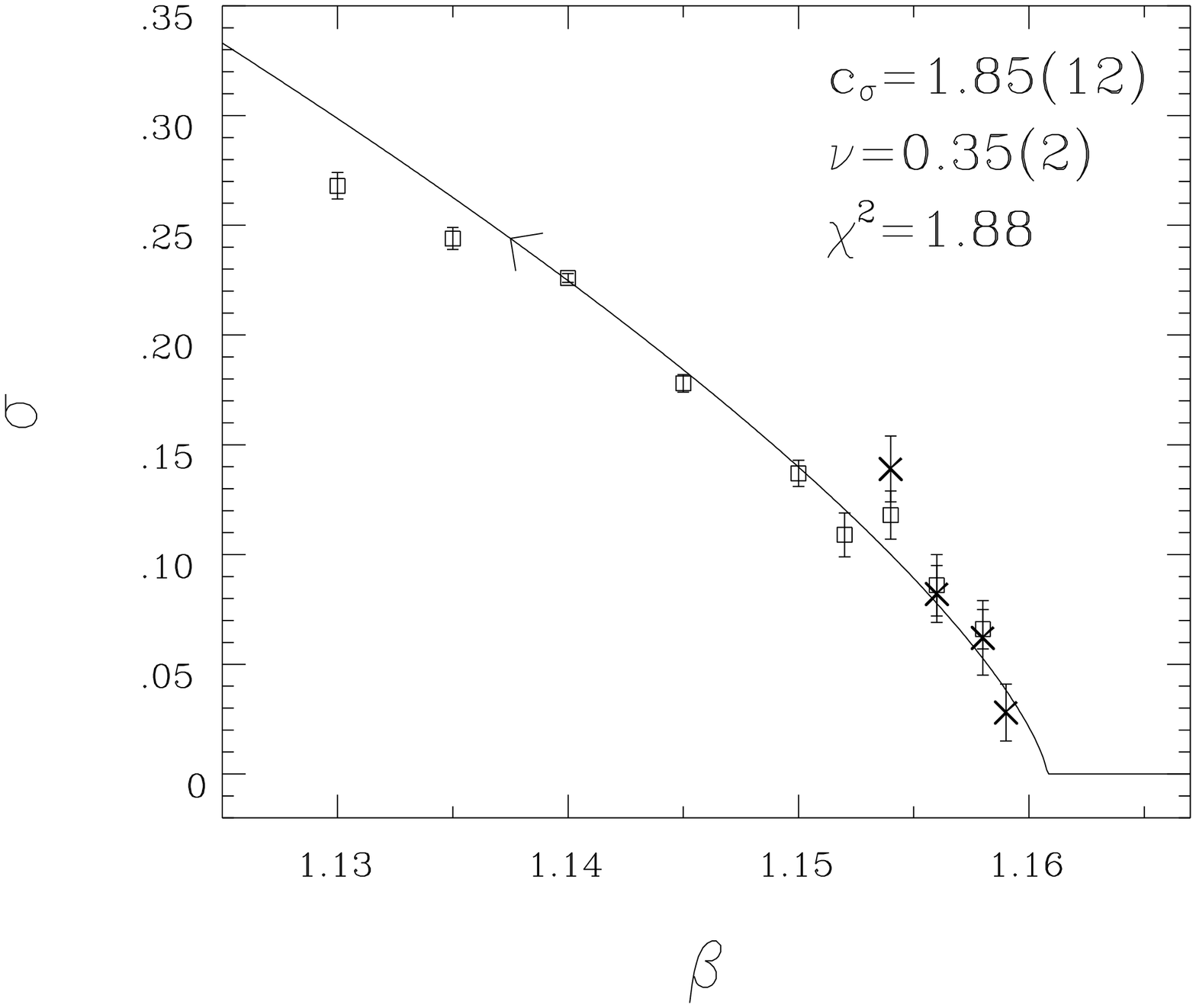,angle=90,width=7cm,bbllx=70,bblly=185,%
        bburx=540,bbury=770}
      \psfig{file=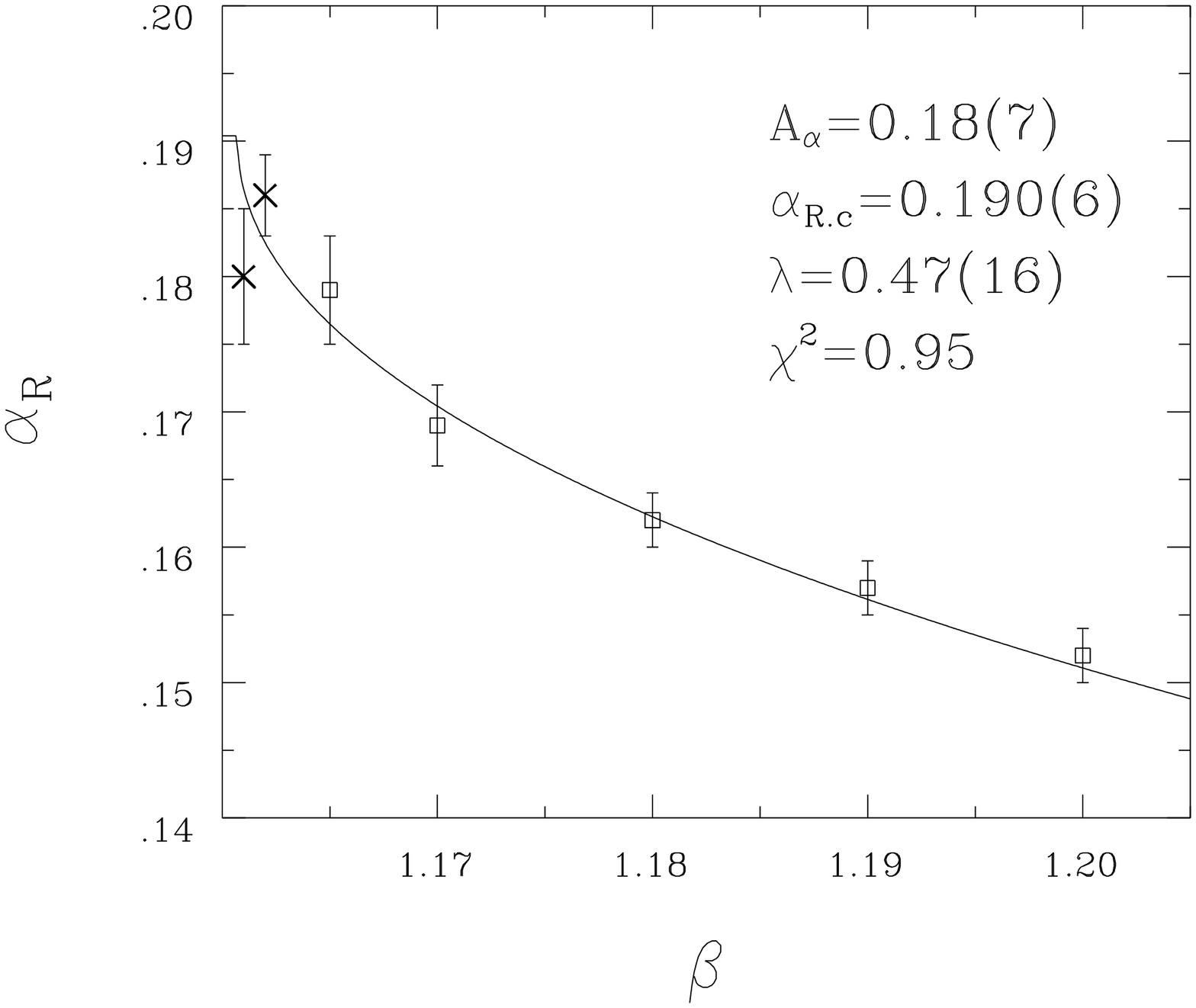,angle=90,width=7cm,bbllx=70,bblly=185,%
        bburx=540,bbury=770}
      }
    }
  \caption{(a) $\sigma$ in the confinement phase. The line is the fit
    \protect{(\ref{SIGMA_SCALING})} of data at $\beta \ge 1.140$. (b) $\alpha_{\rm R}$ in the
    Coulomb phase. The line is the fit
    \protect{(\ref{alfa_scaling})}.  The lattice sizes are $16^3 32$
      (squares) and $20^3 40$ (crosses).}
  \label{fig:alphasigma}
\end{figure}

In order to determine $\alpha_{\rm R,c}$ we extrapolate the values of
$\alpha_{\rm R}$ obtained at noncritical $\beta$ values in the Coulomb
phase to $\beta_c$ via (\ref{alfa_scaling}).  We have used the value
for the critical coupling $\beta_c$, as determined by the scaling of
the gauge ball spectrum.  With three free parameters $\alpha_{\rm
  R,c}, A_{\alpha}$ and $\lambda$ we have obtained
\begin{equation}
      \alpha_{\rm R,c}=0.19(1) \quad,\quad \lambda=0.5(2)
\label{ALPHALAMBDA}
\end{equation}
with $\chi^2/N_{\rm{DF}} = 0.9$. We find that $\alpha_{\rm R,c}$ is
quite well determined by this extrapolation within the Coulomb phase
(cf. fig.~\ref{fig:alphasigma}b).

An alternative determination of the renormalized fine structure
constant at $\beta_c$ can be obtained by assuming that the short range
properties of the static potential are smooth at the critical point.
Thus in the vicinity of the critical point it is sensible to define
$\alpha_{\rm R}$ from the short range part of the static potential in
the confinement phase too; $\alpha_{\rm R,c}$ can then be obtained by
interpolation. This gives results in agreement with
(\ref{ALPHALAMBDA}).

Our result for $\alpha_{\rm R,c}$ at $\gamma = -0.2$ is consistent
with those obtained for other actions \cite{JeNe85} and thus further
supports the conjectured universality of $\alpha_{\rm R,c}$ in the
pure U(1) gauge theory.

\section{The gauge-ball measurements}
\subsection{Quantum numbers}

Our GB operators for a given time-slice are composed of appropriately
symmetrized loops containing links only in the three spatial
directions. The combinations of loops must have a well-defined
behaviour under the point group for cubic symmetry with inversion,
$O_h$.  For rotations one has the proper cubic group $O$ with its five
irreducible representations $R = A_1$, $A_2$, $E$, $T_1$ and $T_2$ of
dimensions 1, 1, 2, 3 and 3 respectively.  Inversion then introduces
the parity quantum number $P=\pm1$, involving direct group product
such that $O_h$ behaves as $O\otimes P$.

As the representations are complex, we can take the real or imaginary
part of the loop, corresponding to charge parity $C=+1$ and $-1$,
respectively.  Thus a complete categorization of a state is $R^{PC}$
with the five possible $R$ as above~\cite{BeBi83}.

In the following we consider all these 20 states with momentum $p=0$
as well as some with $p = 1$ (in units of $2\pi/L_s$).  We denote
these channels by 
\begin{equation}
                j =  R^{PC}(p).
\label{J}
\end{equation}

The representations $R$ contribute in the continuum limit to spin as
indicated in table \ref{repzuord}.  In glue-ball calculations one tends
to identify the lowest energy lattice state with the lowest continuum
spin to which it contributes, but this is not necessarily correct.

\begin{table}[tbp]
  \hfuzz=1pt
  \begin{center}
    \leavevmode
    \footnotesize
    \begin{tabular}{|c|c|c|}\hline
      irr. rep. & dimension & smallest spins\\
      \hline  \hline
      $A_1$ & 1 & 0, 4, ...\\
      \hline  
      $A_2$ & 1 & 3, 6, ...\\
      \hline  
      $E  $ & 2 & 2, 4, ...\\
      \hline  
      $T_1$ & 3 & 1, 3, ...\\
      \hline  
      $T_2$ & 3 & 2, 3, ...\\
      \hline  
    \end{tabular}
  \end{center}
  \caption{The smallest spins that are contained in the irreducible
      representations of the cubic group \protect{\cite{BeBi83}}.
        }
  \label{repzuord}
\end{table}


\begin{table}[tbp]
  \hfuzz=1pt
  \begin{center}
    \leavevmode
    \footnotesize
    \renewcommand{\arraystretch}{1.5}
    \begin{tabular}{|c|c|c|c|}\hline
      type & \#  links & \# orientations & shape\\
      \hline  \hline
      plaquette & 4 & 3 &
      \parbox[c]{3cm}{\makebox[3cm][c]{\rule[-0.2cm]{0cm}{1.0cm}
      \epsfxsize=1.0cm
      \epsfbox{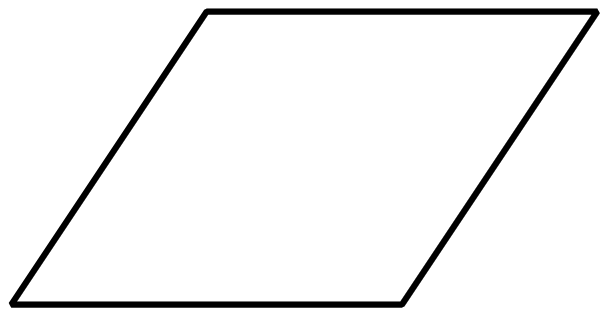}
      }}
      \\
      \hline  
      rectangle & 6 & 6 &
      \parbox[c]{3cm}{\makebox[3cm][c]{\rule[-0.2cm]{0cm}{1.0cm}
      \epsfxsize=1.8cm
      \epsfbox{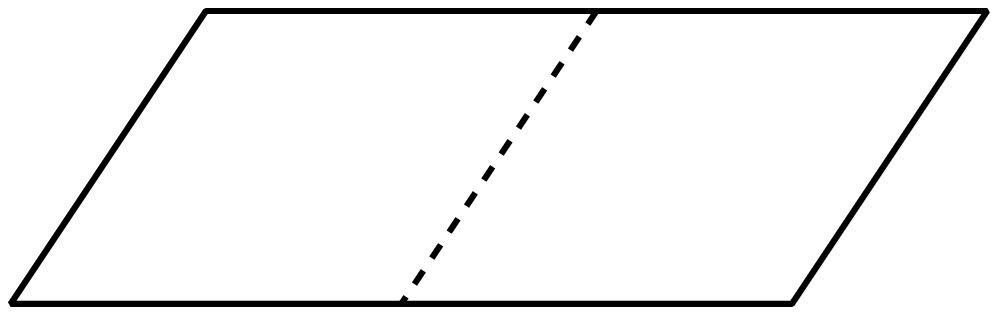}
      }}
      \\
      \hline  
      chair    & 6 &12 &
      \parbox[c]{3cm}{\makebox[3cm][c]{\rule[-0.2cm]{0cm}{1.7cm}
      \epsfxsize=1.0cm
      \epsfbox{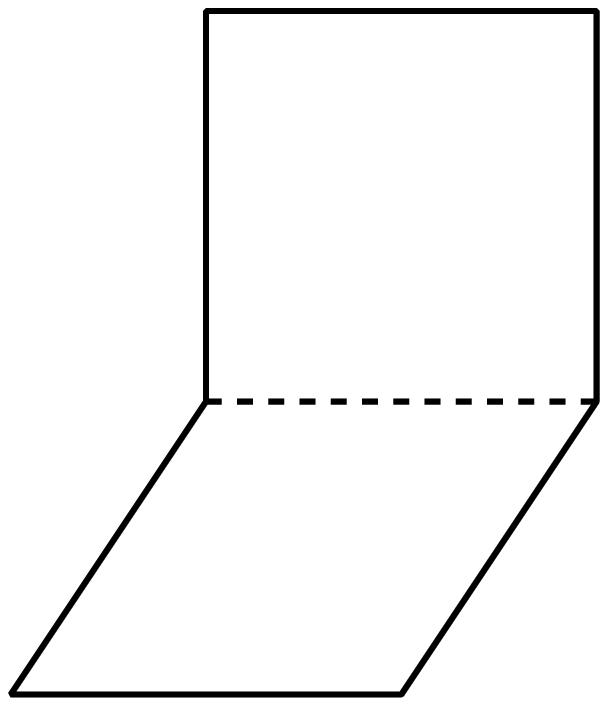}
      }}
      \\
      \hline  
      flange     & 8 &48 &
      \parbox[c]{3cm}{\makebox[3cm][c]{\rule[-0.2cm]{0cm}{1.7cm}
      \epsfxsize=1.8cm
      \epsfbox{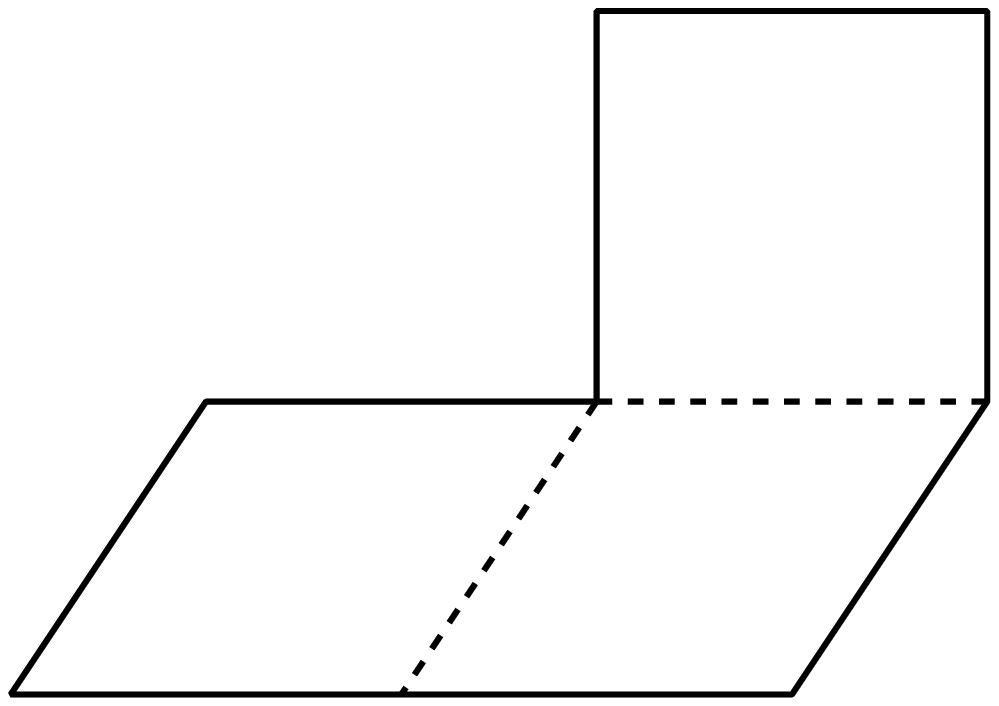}
      }}
      \\
      \hline  
    \end{tabular}
  \end{center}
  \caption
  {Loops used for the construction of the gauge-ball
  operators. Their shapes are extended only in the spatial directions.}
  \label{shapes}
\end{table}


For $p=1$, the states no longer have exactly the symmetry of $O_h$
representations as there is no longer a definite parity.  Thus in
general our non-zero momentum states have been chosen to minimize the
problem of mixing between states. In most channels the $p=1$ results
simply confirm those for $p=0$. In these cases either the mixing is
negligible or the mixed states have higher masses than the ground
state of the `native' channel. Two exceptions in the confinement phase
and one (the photon) in the Coulomb phase are noted in our discussion
of the results.

We have used the loops shown in table \ref{shapes}. In particular the
eight-link operator (flange) has no remaining symmetry and so
contributes to all $J^{PC}$.  One can calculate how the various loop
orientations project onto states of $O_h$ using the tables in
\cite{BeBi83,Mi90}. Thus for zero momentum states we have used
suitable combinations of loops to represent the representations of
$O_h$ and determined the lowest lying states in all channels.  In
order to identify the massless photon in the Coulomb phase we have to
use a non-zero momentum. We discuss this further below.

\subsection{Improved operators and effective energies}

To improve the operators we have used the standard factor-of-two
blocking prescription \cite{Te86Te87}. Accordingly, sets of links on
the original lattice are combined together to form links of twice the
length by adding the contributions from the central link of length
$2a$ plus all four spatial ``staples'' of length $4a$:
\begin{eqnarray}
  U^{(1)}(x,\mu) &=& U(x,\mu)U(x+\hat\mu,\mu) \nonumber\\
  &+& \sum_{{\rm spatial\ } \nu\ne\mu}
  \Big[
  U(x,\nu) U(x\!+\!\hat\nu,\mu) U(x\!+\!\hat\nu\!+\!\hat\mu,\mu)
  U^\dagger (x\!+\!2\mu,\nu) \nonumber\\
  &&+ U^\dagger (x\!-\!\hat\nu, \nu) U(x\!-\!\hat\nu,\mu)
  U(x\!-\!\hat\nu\!+\!\hat\mu,\mu)
  U(x\!-\!\hat\nu\!+\!2\hat\mu, \nu) \Big].
\end{eqnarray}
The resulting complex number is normalized back to an element of U(1)
with unit modulus. This operation is essentially a block-spin
renormalization transformation to a coarser lattice and can be
performed recursively, generating $U^{(2)}$ from $U^{(1)}$ and so on.
The practical limit is reached when the Wilson loop combinations
constructed from these blocked links reach the size of the spatial
lattice.

This gives us many operators $O^r_j(t)$ for a GB state $j$ at
Euclidean time $t$, where the index $r$ now labels all contributions
from different blocking levels and loop shapes to a given $j$. From
this one can form a connected hermitean correlation matrix,
\begin{equation}
  C^{rs}_j(t) = \sum_{t_0} [
   \langle {O^r_j}^*(t_0 + t) O^s_j(t_0)\rangle
   - \langle {O^r_j}^*(t_0 + t)\rangle \langle O^s_j(t_0)\rangle
   ].
\end{equation}

The correlation matrix (at least for a complete set of states) allows
for a spectral decomposition
\begin{equation}
C^{rs}_j(t)= \sum_k A^{rsk}_j\left(e^{-E^k_j t}+e^{-E^k_j(L_t- t)}\right),
\end{equation}
where we assume nondegenerate eigenvalues. The coefficients
$A^{rsk}_j$ represent the projections of our GB operators on the
energy eigenstates.  Since we have only a limited number of states we
have to assume that the diagonalization of the truncated matrix still
gives the physical energies to a good approximation.  The dominant
contribution to the matrix at large $t$ comes from low mass states,
which are the ones we are most interested in.

We now diagonalize the correlation matrix by analogy with
\cite{MiTe89,LuWo90,GaLa93}.  The actual operation we perform can be
written as a generalized eigenvalue problem,
\begin{equation}
C_j(t_0+1) v_j = \lambda_j C_j(t_0) v_j
\label{DIAG}
\end{equation}
in which we pick the eigenvector $v_j$ corresponding to the smallest
energy, and hence the largest eigenvalue $\lambda_j$ (the $rs$ indices
have been suppressed).  Solving this is simplified by the fact that
the $C_j(t)$ have all been symmetrized.  The larger we choose $t_0$
here, the smaller the contribution from higher states. However, the
statistical fluctuations increase rapidly with $t_0$ and in practice
the gain in stability by picking $t_0=0$ is more important when
looking only at the lowest few states.  The corresponding $v_j$
determines the linear combination of operators used for all subsequent
analysis at every $t$.  Its correlator $c_j(t)$ gives the effective
energy and the full correlation matrix is not required again.

Another practical point is that many of the contributions to $C_j$
actually have very similar sets of coefficients $A^{rsk}_j$ for these
lowest states, making the diagonalization less stable.  It should be
remembered that the index $j$ labels blocking level as well as the
Wilson loop shape: we have reduced the number of operators which
appear in the correlation matrix by picking out the loop shape with
the smallest $\epsilon_j(0) = \log (C^{rr}_j(0)/C^{rr}_j(1))$ from
each blocking level, and simply discarding the rest.  Thus between 3
and 5 states actually remain in the correlation matrix, depending on
the lattice size. The effect of the blocking --- the projection onto
the lowest state increases ($\epsilon_j(0)$ decreases) and then falls
off again as the amount of blocking is increased --- means that this
provides a good spread of $A^{rsk}_j$ and the diagonalization in all
cases works well.  Where a comparison is possible the results for the
lowest mass are indistinguishable from those with the full correlation
matrix.  We have found that the blocking and diagonalization of the
correlation matrix works essentially as well as has been found in pure
gauge QCD~\cite{MiTe89}.

The effective energies $\epsilon_j(t)$ are obtained from the
correlation $c_j(t)$ by solving numerically the following two coupled
equations for $\epsilon_j(t)$ and $K_j(t)$,
\begin{eqnarray}
  c_j(t)&=&K_j(t)(e^{-\epsilon_j(t)t}+e^{-\epsilon_j(t)(L_{\rm t}-t)}),\\
  c_j(t+1)&=&K_j(t)(e^{-\epsilon_j(t)(t+1)}+e^{-\epsilon_j(t)
    (L_{\rm t}-t-1)}).\nonumber
\end{eqnarray}
Their approximate solution for small $t$ is
\begin{equation}
      \epsilon_j(t)=\log\frac{c_j(t)}{c_j(t+1)}.
\end{equation}

\subsection{Errors of effective energies}

The errors of the effective energies $\epsilon_j(t)$ were obtained
directly by the bootstrap method.  In this procedure, data consisting
of $N_d$ sets is sampled to produce a new ensemble of $N_d$ sets; each
sample is completely random, so that the new ensemble can contain any
of the original sets more than once or not at all.  This ensemble is
then processed in exactly the same way as the original.  The procedure
is then repeated an arbitrary but suitably large number of times
$N_{bs}$, producing a spread of $N_{bs}$ results.  The standard
deviation of these results gives a reliable estimate of the
statistical error in the original ensemble.

Our initial data was blocked down by a factor 50, producing an $N_d$
of typically a few tens of data sets for each observable.  We have
picked $N_{bs} = 99$.  We have checked that using the same variational
basis for all the bootstrap data, rather than re-diagonalizing the
correlation matrix for each bootstrap sample, does not affect the
errors obtained, and have therefore performed the bootstrap on the
data only after the diagonalization.

We have arbitrarily chosen only to look at the first eight
correlations and hence seven $\epsilon_j(t)$ for $t=0$ to 6.  The
choice of the range of $t$ over which we perform fits to extract the
real energies $E_j$ is described in the next section.

We have analyzed the appropriate range of effective energies in two
ways: firstly, using a one-parameter uncorrelated fit to the data over
the chosen range of $t$, and secondly by taking a weighted mean of all
the values over the range.  Using the second method we processed the
bootstrap samples with the same weights, giving an estimate of the
overall statistical error which is presumably more reliable than that
on the fit parameter.  Certainly the errors with this method appear
consistently larger than with the other; some effective energies have
an error larger by a factor of around two.  This is particularly
noticeable in the states near to the phase transition with a good
signal.  We have therefore used the values from the bootstrap of the
weighted means for our statistical errors.

This analysis gives essentially statistical errors only. An estimate
of the systematic errors of our final results will be made at the end
of the next section.

\section{Scaling of the GB masses in the confinement phase}

\subsection{Gauge-ball energies and masses}

The list of GB channels we have investigated is given in the first
column of table \ref{final_spectrum}. The momenta considered are
indicated in parentheses. In the second column we give the quantum
numbers of the expected continuum state with the smallest spin, in
accordance with the table \ref{repzuord}.

\begin{table}
  \caption{GB states with momentum $p=0, 1$ (in the units $L_s/2\pi$) observed
    in various channels in the confinement phase and their scaling
    behaviour. If two momenta are listed in one line, the mass has
    been assumed to be the same. The quantum numbers of the continuum
    states with smallest possible spin are indicated in the column
    ``continuum''. The clarity of the effective energy plateau and the
    reliability of the mass determination is indicated by a
    corresponding number of ``+''. In the channels denoted ``?''  the
    evidence for a mass is not clear.  In the columns ``$\nu$''
    and ``mass'' the scaling exponent of the GB mass and its
    approximate proportionality to one of the mass scales is
    indicated. The ratio of each mass to the corresponding mass scale
    is given in the column ``ratio''. For the explanation of the
    continuum states marked with $^*$ see the text.  }

  \label{final_spectrum} 
  \begin{center}
    \leavevmode
    \footnotesize
    \begin{tabular}{|l|l|c|c|c|c|}\hline
      $R^{PC}(p)$&continuum & quality&$\nu$ & mass &ratio  \\
      \hline      
      \hline      
      $A_1^{++}(0,1)$   &$0^{++}$ & & & &1  \\       
      $T_1^{-+}(1)\;^*$ &$0^{++}\: ^*$
      &&\raisebox{1.5ex}[-1.5ex]{$\nu_{\rm g}$} 
      &\raisebox{1.5ex}[-1.5ex]{$m_{\rm g}$}  & 1.01(6)\\ 
      \cline{1-2}\cline{4-6}      
      $A_2^{+-}(0)$     &$3^{+-}$  & & & &1.00(4) \\ 
      $E^{+-}(0)$       &$2^{+-}$ &$+\!+\!+$  & & & 0.98(4)\\ 
      $T_1^{+-}(0,1)$   &$1^{+-}$ & &&$m_{\rm ng} $  & 1 \\ 
      $T_2^{+-}(0,1)$   &$2^{+-}$ & & &  & 1.00(4)\\ 
      $T_2^{--}(1)\;^*$&$2^{+-}\: ^*$ & & &  & 1.04(5)\\ 
      \cline{1-3}\cline{5-5}      
      $E^{++}(0,1)$ &$2^{++}$ & & & & 2.06(9)\\ 
      $T_2^{++}(0,1)$&$2^{++}$ & \raisebox{1.5ex}[-1.5ex]{$++$}& &\raisebox{1.5ex}[-1.5ex]{$\simeq 
             2m_{\rm ng}$}   & 2.07(9)\\
      \cline{1-3}\cline{5-5}      
      $A_1^{-+}(0)$ &$0^{-+}$ & & &  & 2.88(13)\\ 
      $A_1^{--}(0)$ &$0^{--}$ & &$\nu_{\rm ng}$  &  & 2.73(14)\\ 
      $A_2^{++}(0)$ &$3^{++}$ & & &  & 3.22(16)\\ 
      $A_2^{--}(0)$ &$3^{--}$ & & &  & 2.73(15)\\ 
      $E^{--}(0)$ &$2^{--}$ & & & & 2.95(14)\\ 
      $T_1^{++}(0,1)$ &$1^{++}$ &\raisebox{1.5ex}[-1.5ex]{$+$}& &\raisebox{1.5ex}[-1.5ex]{$(2.6 -  
           3.6)m_{\rm ng}$}    & 3.61(18)\\ 
      $T_1^{--}(0,1)$ &$1^{--}$ & & & & 2.94(13)\\ 
      $T_2^{-+}(0)$ &$2^{-+}$ & & & & 3.10(14)\\ 
      $T_2^{-+}(1)\;^*$ &$2^{++}\:^*$ & & & & 2.63(12)\\ 
      $T_2^{--}(0)$ &$2^{--}$ & & & & 3.25(16)\\ 
      \cline{1-3}\cline{5-5}      
      $\sqrt{\sigma}$ & & & & $1/3 m_{\rm ng}$& 0.34(2)\\ 
      \hline
      $A_1^{+-}(0)$ &$0^{+-}$ && \multicolumn{3}{c|}{}   \\ 
      $A_2^{-+}(0)$ &$3^{-+}$ && \multicolumn{3}{c|}{}   \\ 
      $E^{-+}(0)$   &$2^{-+}$ & \raisebox{1.5ex}[-1.5ex]{?}&
      \multicolumn{3}{c|}{\raisebox{1.5ex}[-1.5ex]{?}} \\ 
      $T_1^{-+}(0)$ &$1^{-+}$ && \multicolumn{3}{c|}{}   \\
      \hline      
    \end{tabular}
  \end{center}
\end{table}


In two cases our results for the confined phase should be interpreted
as showing a clear signal for mixing when the momentum is non-zero, and
hence the naive $J^{PC}$ interpretation is not correct.  These states,
the $T_1^{-+}$ and $T_2^{--}$, are marked with an asterisk in table
\ref{final_spectrum}.  For $p=0$ they give no clear signal, increasing
the likelihood that states mixed from other $J^{PC}$ will be visible.
To see that the mixing is as expected, one needs to consider the
behaviour of $O_h$ states when the symmetry is broken along a
particular axis, as by our $(1,0,0)\times2\pi/L_s$ momentum boost.
The reduced symmetry group is dihedral: one finds that the $T_1^{-+}$
component whose own axis of symmetry lies along the momentum axis
(i.e. longitudinally polarized) has the same behaviour under the
reduced symmetry group as the $A_1^{++}$, and that the $T_2^{--}$
likewise mixes with the $T_2^{+-}$.  This agrees with our results.
Also the channel $T_2^{-+}(1)$ might have an admixture from one of the
$2^{++}$ states (see below).

In each channel we have determined the effective energies
$\epsilon_j(t)$ for as many $t \le 6$ as possible. In principle
one should obtain each GB energy $E_j$ by a fit to their plateau which
is expected at large $t$. In practice, in some channels plateaus can
be found only at moderate $t$, since $\epsilon_j(t)$ has very large
errors or is unmeasurable for larger $t$. The quality of the plateaus
is therefore different in different channels and varies with $\beta$,
being best close to $\beta_c$.  Sometimes we cannot identify a plateau at all.

We therefore classify the channels according to the quality of the
signal for a definite GB energy $E_j$, and extract this energy in
somewhat different ways.  The classification is indicated in the third
column of table \ref{final_spectrum}; this holds only for the
confinement phase.  The meaning is as follows:

\begin{description}

\item [\bf $+++$ states:] At most $\beta$ a plateau in $\epsilon_j(t)$
  was found and fitted at $t \ge 2$.

\item [\bf $++$ states:] At most $\beta$ a plateau in $\epsilon_j(t)$
  was found only if $t = 1$ was included. This plateau was then fitted.

\item [\bf $+$ states:] A plateau in $\epsilon_j(t)$ was found only at
  the highest statistics point $\beta = 1.159$ if $t = 1$ was
  included. This plateau was then fitted. For other $\beta$ the value
  of $\epsilon_j(1)$ was taken for the GB energy $E_j$.

\item [\bf $?$ states:] No plateau in $\epsilon_j(t)$ was found at $t
  \ge 1$ even for $\beta = 1.159$. Then we do not obtain any GB energy.

\end{description}

\begin{figure}[htb]
  \centerline{
    \hbox{
      \psfig{file=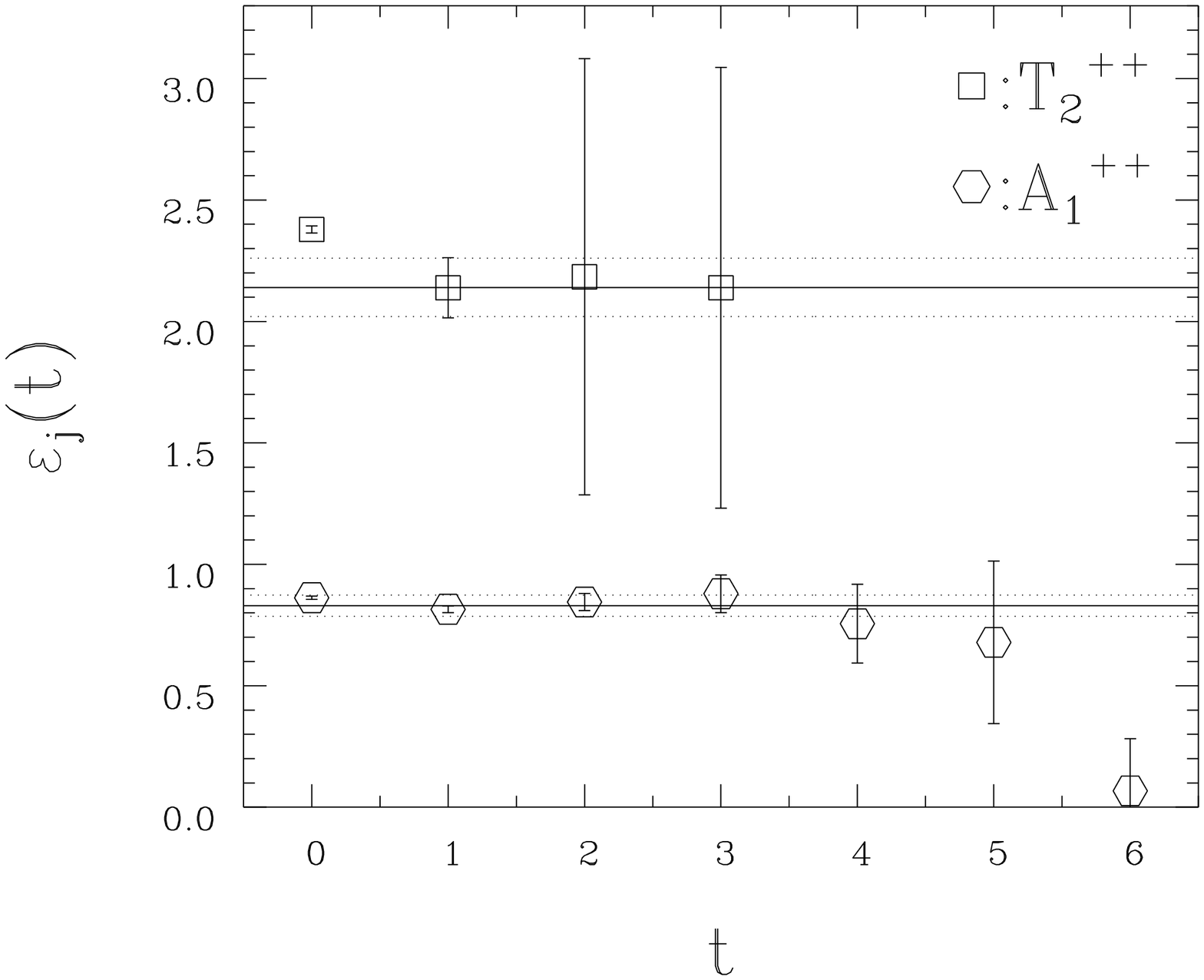,angle=90,width=7cm,bbllx=70,bblly=185,
        bburx=540,bbury=770}
      \psfig{file=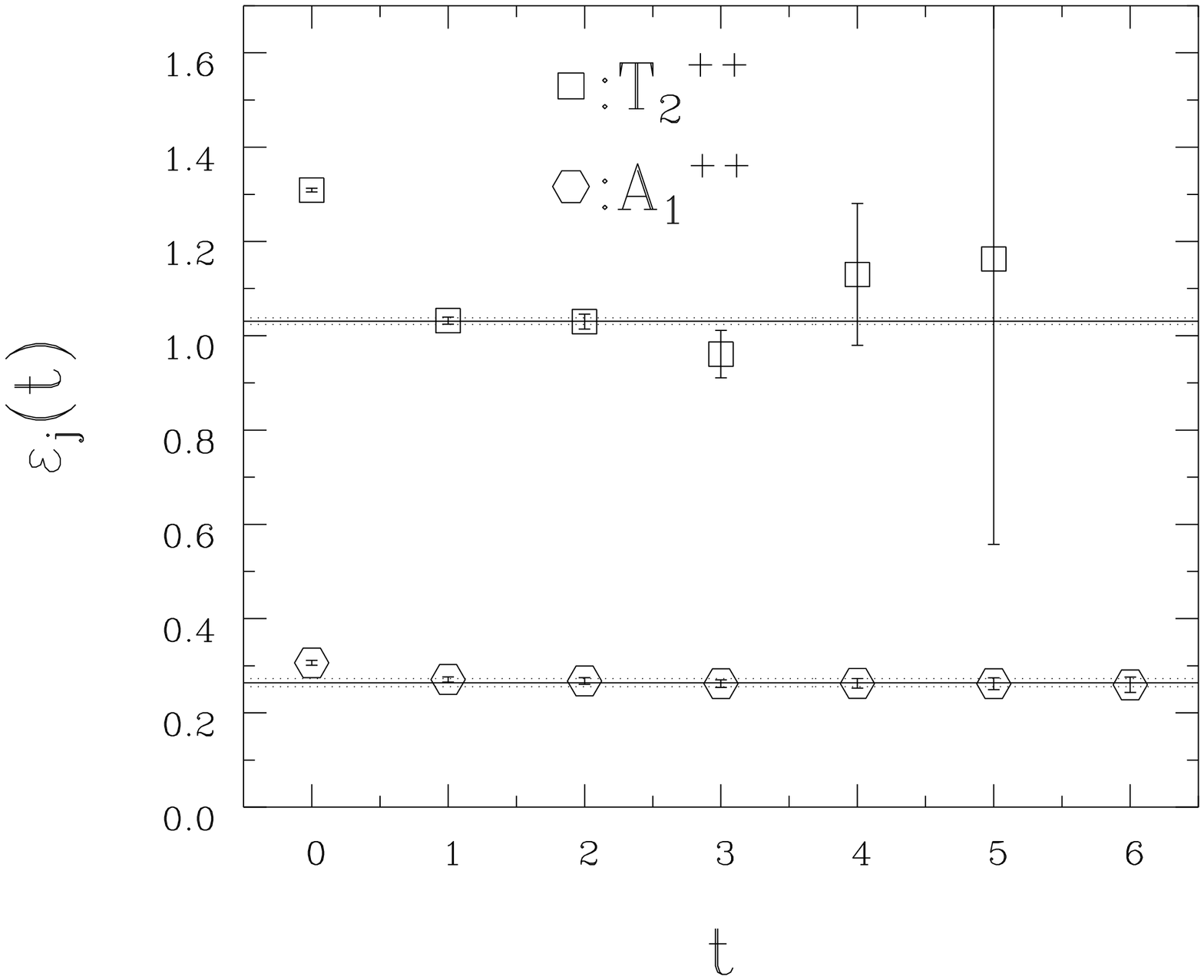,angle=90,width=7cm,bbllx=70,bblly=185,
        bburx=540,bbury=770}
      }
    }
  \caption{Effective energies from $A_1^{++}(0)$ and $T_2^{++}(0)$ correlation
    functions (a) at $\beta=1.145$ on the $16^3 32$ lattice and (b) at
    $\beta=1.159$ on the $20^3 40$ lattice. The plateaus were fitted
    omitting the points at $t=0, 1$ for $A_1^{++}$ and the point $t=0$
    for $T_2^{++}$. The GB energies $E_j$ and their errors are
    indicated by the horizontal lines.}
  \label{fig:effmass}
\end{figure}

In order to illustrate the quality of the data, we show in
fig.~\ref{fig:effmass}a results for the effective energies of
$A_1^{++}$ ($+\!+\!+$ quality) and $T_2^{++}$ ($++$ quality) with
$p=0$ obtained at a typical data point $\beta = 1.145$ on the $16^3
32$ lattice.  The effective energies of the same states at the data
point with best statistics and smallest errors, $\beta = 1.159$ (close
to $\beta_c$) on the $20^3 40$ lattice, are shown in
fig.~\ref{fig:effmass}b. The GB energies $E_j$ with errors are shown
as horizontal lines.

The GB energies $E_j$ at $p = 0$ and $p = 1$ have been used to
calculate the GB masses $m_j$ by the lattice dispersion relation
\begin{equation}
  2(\cosh E_j - 1) = m_j^2+2\sum_{\mu=1}^3 (1-\cos p_\mu).
\label{DISP_REL}
\end{equation}
If the masses found in the $p=0$ and $p=1$ channels with the same
$R^{PC}$ turned out to be consistent, we have continued the analysis
assuming their equality, i.e. they were fitted by exactly the same
parameters. In the table \ref{final_spectrum} these ``pair'' channels
are represented by one line only, as well as by the same symbol in the
figures \ref{fig:nu} and \ref{fig:spectrum}.

\subsection{Two mass scales}

Having determined the GB masses for various $\beta$ we have
investigated their scaling behaviour with $\beta$. First the fits of
the form
\begin{equation}
                    m_j = c_j (\beta_c^j - \beta)^{\nu_j}
\label{FIT_INDIVIDUAL}
\end{equation}
were performed for each GB channel $j$ individually. We found two
groups of masses with strikingly different scaling behaviour. Within the
whole $\beta$ range a large group of the GB masses scales with roughly
the same exponents $\nu_j$ close to the non-Gaussian value 0.365(8)
found in \cite{JeLa96a,JeLa96b,LaPe96}. However, in a small group
consisting only of the $A_1^{++}$, $p=0,1$ and the $T_1^{-+}$, $p=1$
channels the values of $\nu_j$ are approximately Gaussian, i.e. $1/2$.
The values $\beta_c^j$ are quite consistent with each other in all
channels. Therefore, in the further analysis we have assumed the same
value of $\beta_c$ in all channels.

\begin{figure}[tbp]
  \begin{center}
    \psfig{file=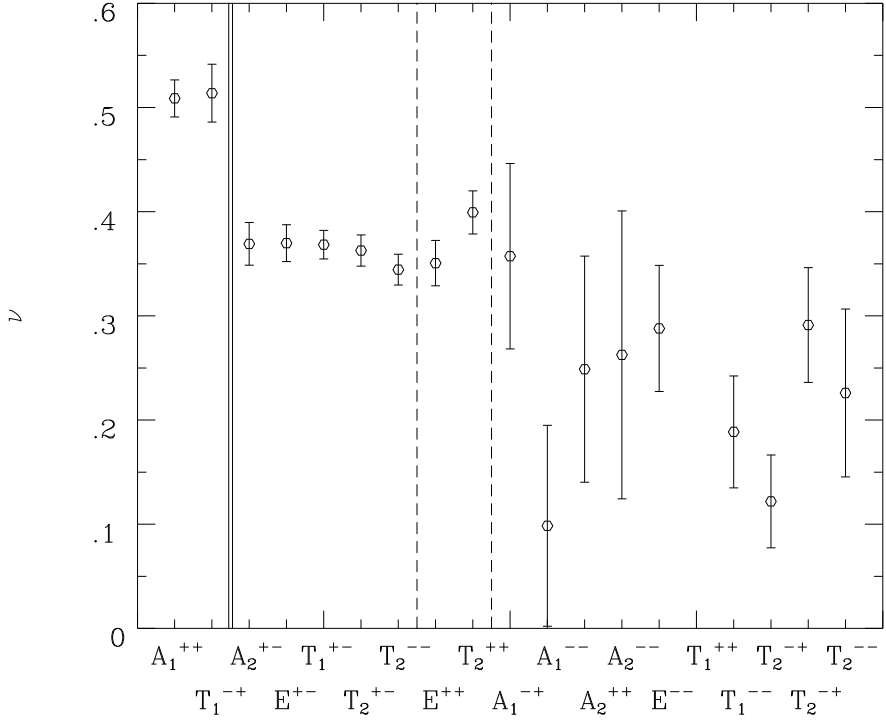,angle=90,width=\hsize,bbllx=50,
      bblly=110,bburx=545,bbury=774}
  \end{center}
  \caption{Values of $\nu_j$ obtained in each channel separately. The
    double vertical line separates two groups with distinctly
    different $\nu_j$. The dashed vertical lines separate channels
    with different quality signal within the group with non-Gaussian
    exponent values. The ordering within this group is with decreasing
    reliability (table \ref{final_spectrum}) from left to right.}
  \label{fig:nu}
\end{figure}

\begin{figure}[tbp]
  \begin{center}
    \psfig{file=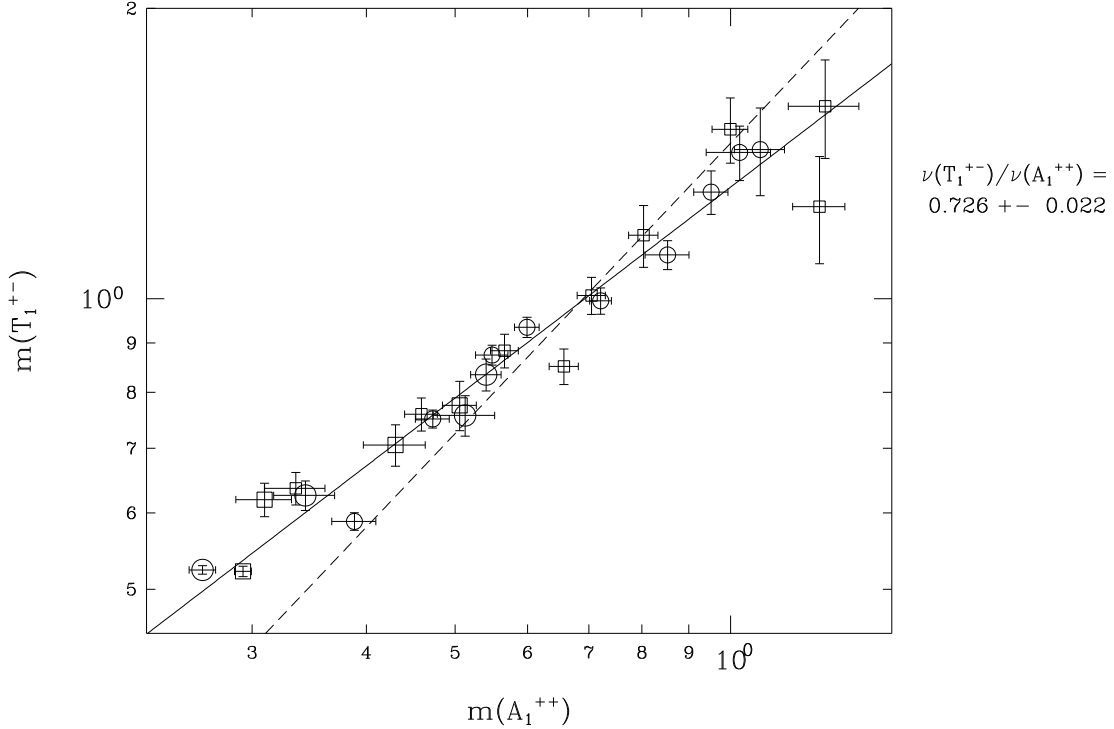,angle=90,width=\hsize,bbllx=50,
      bblly=110,bburx=545,bbury=774}
  \end{center}
  \caption{A log-log plot of the masses of the $A_1^{++}$
    and $T_1^{-+}$ gauge-balls in the interval $1.13\leq \beta\leq
    1.159$. The results from the channels with $p=0,1$ are denoted by
    circles and squares, respectively. The solid line is the result of
    a straight line fit considering errors in both directions. The
    dashed straight line with a gradient of one, i.e. assuming equal
    exponents, is shown for comparison. }
  \label{fig:loglog}
\end{figure}

The results for $\nu_j$ obtained from (\ref{FIT_INDIVIDUAL}) with
common $\beta_c$ are shown in fig.~\ref{fig:nu}. A clustering around
two values is clearly seen. In order to further demonstrate the
difference of the scaling behaviour between the two groups we plot the
masses of the clearest members of each group, the $T_1^{+-}$ and
$A_1^{++}$ channels with $p=0,1$, in a log-log plot against each other
in fig.~\ref{fig:loglog}.  The data have a slope distinctly different
from one.

\begin{figure}[tbp]
  \begin{center}
    \leavevmode
    \psfig{file=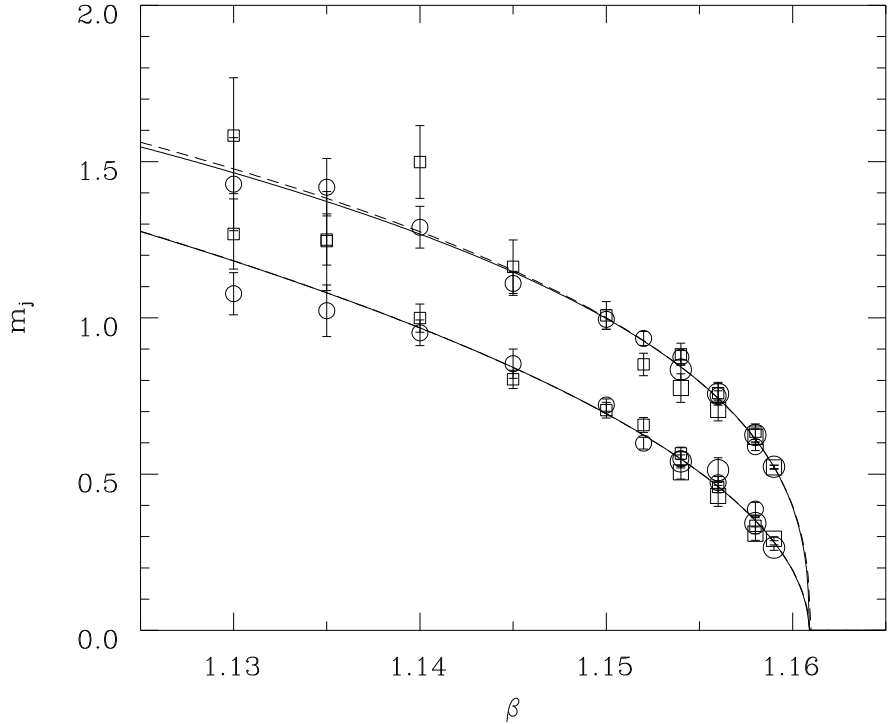,angle=90,width=\hsize,bbllx=50,
      bblly=110,bburx=545,bbury=774}
  \end{center}
  \caption{Masses of the $A_1^{++}$ (lower points) and $T_1^{+-}$
    glueballs in the confinement phase versus $\beta$ for both
    momenta ($p=0,1$ being denoted by circles and squares,
    respectively). The dashed curves are fits to the data shown.
   The full curves represent the mass scales
    \protect{(\ref{MSCALES})}. }
  \label{fig:2scale}
\end{figure}

Naturally, as seen in fig.~\ref{fig:nu}, the errors of $\nu_j$ in
channels with less accurately determined masses are rather large.
However, for example in the $T_1^{+-}$ and $A_1^{++}$ channels with
$p=0$ and $p=1$ the masses are quite accurate and the statistical
errors are small. Performing fits to each of these two channels with
both momenta we obtain
\begin{eqnarray}
                \nu_{(T_1^{+-},p=0,1)} &=&  0.37(3),   \nonumber \\
                \nu_{(A_1^{++},p=0,1)} &=&  0.51(3).
\label{NUINDIV}
\end{eqnarray}
The masses in these two channels with both momenta (distinguished by
different symbols) are shown in fig.~\ref{fig:2scale}. The
corresponding fits by means of (\ref{FIT_INDIVIDUAL}) are indicated by
dashed lines.

These observations suggested the next step of the analysis: we have
assumed the same exponent $\nu_j$ for each group, denoted $\nu_{\rm
  ng}$ and $\nu_{\rm g}$ for the non-Gaussian and Gaussian group,
respectively. Then a joint fit was performed with these two exponents,
common $\beta_c$, and the individual amplitudes $c_j$ as free
parameters:
\begin{equation}
                    m_j = c_j \tau^{\nu_{\rm f}}, \quad {\rm f = ng, g}.
\label{FIT_GLOBAL}
\end{equation}

The resulting values of the exponents are
\begin{eqnarray}
                \nu_{\rm ng} &=&  0.367(14),             \nonumber \\
                \nu_{\rm g}  &=&  0.51(3).
\label{NUGLOB}
\end{eqnarray}
They are essentially determined by the most accurately determined
masses, such as those in the $T_1^{+-}(0,1)$ and $A_1^{++}(0,1)$
channels. The critical point at $\gamma = -0.2$ determined in this way
is
\begin{equation}
               \beta_c = 1.1609(2).
\label{BETA_C_STAT}
\end{equation}
The amplitudes, giving the ratios of the masses in each group, are
described below.

All the reliably determined GB masses are well described by this fit.
Those which are not fully reliable are at least consistent with it.
{}From this we conclude that the system has two mass scales which we
denote by $m_{\rm ng}$ and $m_{\rm g}$. Each GB mass scales according
to one of these mass scales. We choose the fitted scaling behaviour of
the GB operators $T_1^{+-}$ and $A_1^{++}$ to define the two mass
scales:
\begin{eqnarray}
  m_{\rm ng} &=& c_{\rm ng} \tau^{\nu_{\rm ng}},
  \quad c_{\rm ng} =  c_{(T_1^{+-}, p=0,1)} = 5.4(3), \nonumber\\
  m_{\rm g} &=& c_{\rm g} \tau^{\nu_{\rm g}},
  \quad c_{\rm g}  =  c_{(A_1^{++}, p=0,1)} =  7.4(6).
\label{MSCALES}
\end{eqnarray}

The full curves in fig.~\ref{fig:2scale} represent these mass scales
$m_{\rm ng}$ and $m_{\rm g}$. We observe that the dashed curves are
almost indistinguishable from the mass scales (\ref{MSCALES}). Our
determination of these scales is thus not sensitive to the selection
of only one channel (with both $p$) data for their definition, nor to
the use of a joint value for $\beta_c$.

\subsection{Gauge-ball spectrum}

Table~\ref{final_spectrum} summarizes the results for the GB masses in
the confinement phase and gives the observed masses in multiples of
the two ``standard'' masses $m_{\rm ng}$ and $m_{\rm g}$.  For the
states denoted by $+$ our assignment is only tentative.

\begin{figure}[tbp]
  \begin{center}
    \psfig{file=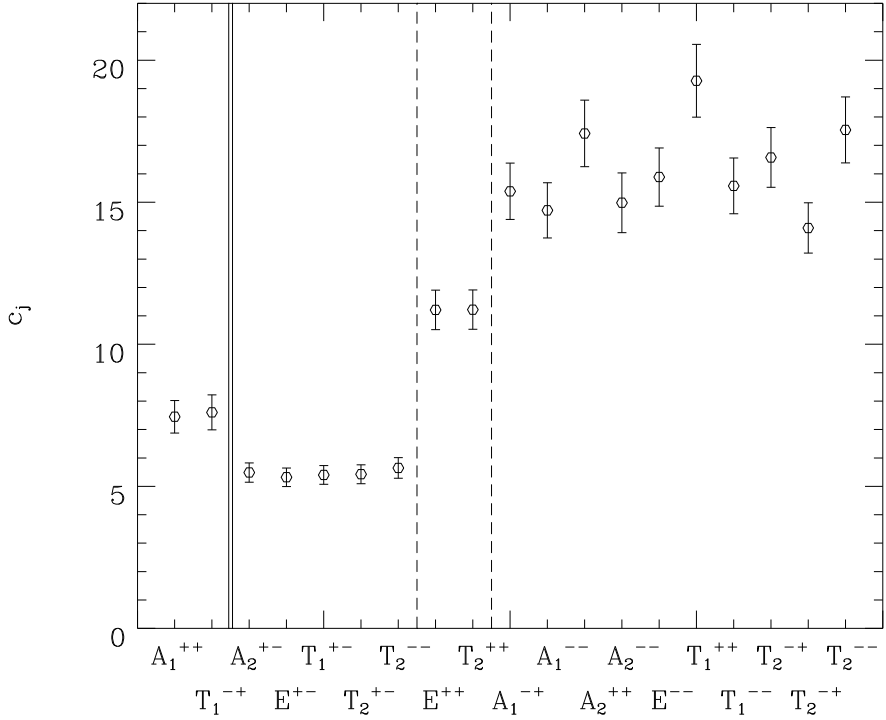,angle=90,width=\hsize,bbllx=50,
      bblly=110,bburx=545,bbury=774}
  \end{center}
  \caption{The amplitudes $c_j$ of the fits
    \protect{(\ref{FIT_GLOBAL})}.  The vertical lines have the same
    meaning as in \protect{fig.~\ref{fig:nu}}.}
  \label{fig:spectrum}
\end{figure}

The results for the individual masses are presented in
fig.~\ref{fig:spectrum} and in the last two columns of
table~\ref{final_spectrum}. In fig.~\ref{fig:spectrum} the individual
amplitudes $c_j$ are shown.  They summarize our results for the masses
obtained from all $\beta$ points in the confinement phase and reflect
their scaling behaviour.  It is apparent that the states in the
non-Gaussian group cluster around three values of $c_j$, being
multiples of the lowest mass states in this group.

\begin{figure}[tbp]
  \begin{center}
    \leavevmode

\psfig{file=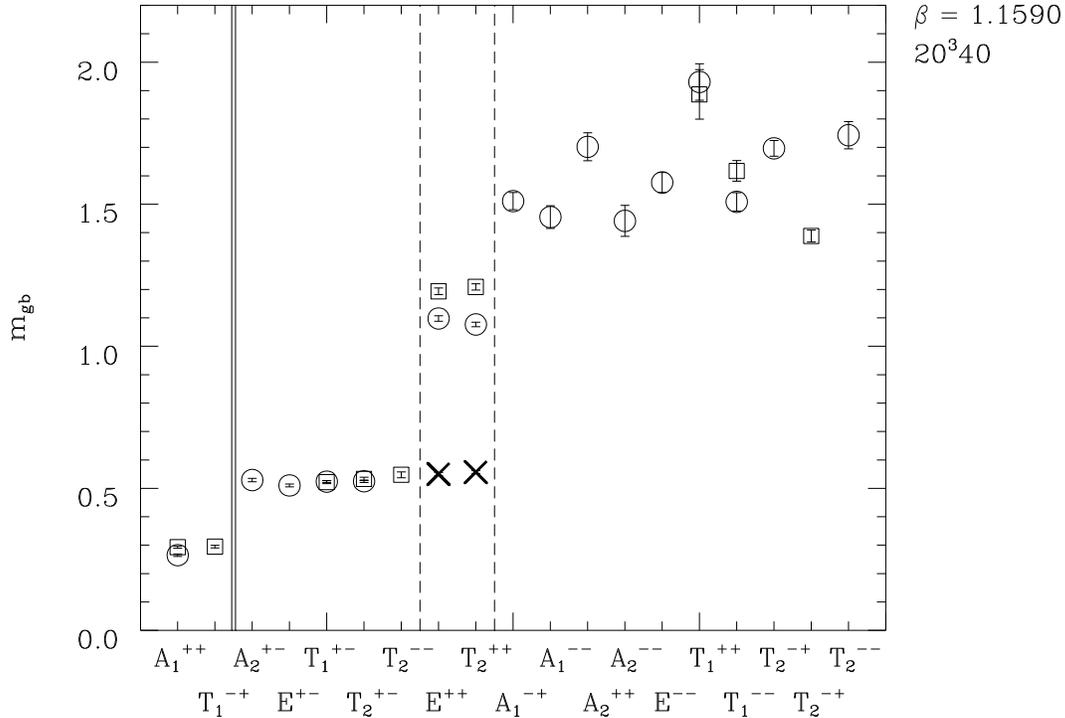,angle=90,width=\hsize,%
       bbllx=50,bblly=110,bburx=545,bbury=774}
  \end{center}
  \caption{GB masses $m_j$ at the point with the highest statistics, $\beta =
    1.159$ on the $20^3 40$ lattice. The $p=0,1$ results are indicated
    by the circles and squares, respectively. The crosses belonging to
    the $E^{++}(0,1)$ and $T_2^{++}(0,1)$ channels are masses of
    individual particles when the energies $E_j$ in these channels 
    with $p=1$ are assumed to correspond to two-particle states with
    lowest nonzero momentum. The vertical lines have the same meaning
    as in \protect{fig.~\ref{fig:nu}}.}
  \label{fig:spectrum_1159}
\end{figure}

An alternative way to determine the GB masses is to rely only on the
high statistics data point, $\beta = 1.159$ on the $20^3 40$ lattice.
The absolute values of $m_j$ at this point are presented in
fig.~\ref{fig:spectrum_1159}. Here we show the results for $p=0$ and
$p=1$ separately in all channels, including the pairs. Of course, at
one fixed $\beta$ the difference in the scaling behaviour is not
observable. However, keeping in mind that there are two groups of
states, one can compare figs.~\ref{fig:spectrum} and
\ref{fig:spectrum_1159} for each group separately and find an
agreement between the mass ratios. The two-fold and perhaps three-fold
clustering of the values in the non-Gaussian group around the lowest
mass is again apparent.

As seen in fig.~\ref{fig:spectrum_1159}, the dispersion relation
(\ref{DISP_REL}) is slightly violated for the pairs $E^{++}(0,1)$ and
$T_2^{++}(0,1)$. Motivated by the fact that the corresponding masses
are very close to $2m_{\rm ng}$, we have also tried to apply the
dispersion relation assuming that it is really two-particle states
which are seen in these channels, either both particles having zero
momentum or one of the particles moving with the lowest nonzero
momentum.  Indeed, the fit of $E_j$ in these channels gave the mass
$m_{\rm ng}$ for each of the two particles even for $p=1$. This result
is indicated by crosses in fig.~\ref{fig:spectrum_1159}. It suggests
that the $E^{++}(0,1)$ and $T_2^{++}(0,1)$ channels are dominated by
states of two gauge-balls from the non-Gaussian group with mass
$m_{\rm ng}$. Of course, further verification of such a possibility is
necessary.

Only the ratios of the amplitudes within each group are relevant for
the continuum limit, and therefore we show in the last column of
table~\ref{final_spectrum} the ratios $r_j$ to the corresponding mass
scales (\ref{MSCALES}),
\begin{equation}
          m_j = r_j m_{\rm f},  \quad {\rm f = ng, g}.
\label{ratios}
\end{equation}

Relating the different scales set by the string tension and by the
mass $m_{ng}$
\begin{equation}
        \sqrt{\sigma}=r_{\sigma}m_{ng},
\label{SIGMANG}
\end{equation}
we obtain
\begin{equation}
        r_{\sigma}=0.34(2).
\end{equation}\label{RATSIGMANG}
This value is given in table~\ref{final_spectrum} for a comparison
with the GB masses.

It is interesting to compare our results in the confined phase with
those of quenched QCD (i.e. pure gauge SU(3)), for which the most
recent analysis of all $J^{PC}$ appeared in ref.~\cite{BaSc93} (see in
particular fig. 1), and of the $A_1^{++}$ and $E^{++}$ also in
\cite{ChSe94}.  Although our results here are spread over a larger
variety of lattice sizes and couplings, the higher efficiency of U(1)
calculations implies that the statistical accuracy of the results is
comparable in the two cases.

As in SU(3), we find that masses tend to appear together in bands.
This is more pronounced in our case as it appears in low-lying and
therefore well measured states.  In both cases we find the lowest
lying state to be the most symmetric, namely $0^{++}$.  We also find a
low-lying $2^{++}$ (lattice $E^{++}$ and $T_2^{++}$) state.  A clear
difference to SU(3) is that in the U(1) case there is a band of
intervening $PC={+-}$ states for spins 1, 2 and 3, all with similar
mass, which do not appear in SU(3) (though in the latter case there is
a $1^{+-}$ state with approximately twice the lowest glueball mass).
Indeed, these states in U(1) are the lowest ones with the non-Gaussian
scaling behaviour.  On the other hand, we do not find the low-lying
$0^{-+}$ (and maybe $0^{+-}$, which had a large error) of SU(3).

Heavier states have correspondingly large errors, making any further
comparison too difficult.  It is at least clear that the similarity of
confining behaviour is not by itself enough to guarantee a similar
spectrum.

One further similarity to QCD is remarkable: It has been known for
long that in the lattice QCD the $\beta$ dependence of the $0^{++}$
glueball is somewhat different from that of the other glueball
states. We shall return to this point in the next section.

The simplest interpretation of the observed spectrum might be the
following: The Gaussian spectrum consist of one scalar $0^{++}$ only.
The non-Gaussian group has a nontrivial but quite simple spectrum
consisting of the $1^{+-}$, $2^{+-}$ and $3^{+-}$ states of the same
mass $m_{\rm ng}$. Higher spin assignments for the $J^{+-}$ states are
not ruled out, however. In the other channels the masses are $\simeq
2m_{\rm ng}$ and $(2.6 - 3.6)m_{\rm ng}$. They may well correspond to
two- and three-particle states of the lightest non-Gaussian
gauge-balls. The multi-particle states of the Gaussian scalar do not
seem to contribute significantly to the non-Gaussian channels in spite
of its low mass. This further supports the expectation that the
$0^{++}$ belongs to a ``trivial'' theory.

\subsection{Error analysis}

The errors given up to now in this section have been obtained by
purely statistical procedures. They are based on the essentially
statistical errors of the energies $E_j$ and the corresponding masses
$m_j$, described in the previous section. As seen in (\ref{NUGLOB})
and (\ref{BETA_C_STAT}), they are rather low, on the level of 4\% for
$\nu_{\rm ng}$.

However, it is obvious that various additional errors -- partly of
systematic origin -- should be expected. The main reason is that we
have to stay away from the critical point in order to avoid finite
size effects and the two-state signal. The value of $\beta_c$ then
enters very sensitively into the determination of $\nu$ by fits like
(\ref{FIT_GLOBAL}). Furthermore, mass values $m_j \simeq 1$ have to be
used in such a fit. This raises the question of nonleading terms in
the scaling behaviour.

To get at least an estimate of these effects, we made several analyses
of the scaling behaviour of the well determined ($+++$) masses when
$\beta_c$ is approached, modifying the procedure in various ways. In
most cases the obtained values for $\nu$ and $\beta_c$ varied slightly
with the procedure.
\begin{enumerate}
\item We modified the $\beta$ interval in which the power law fits
  were performed, dropping up to five $\beta$ points most distant from
  $\beta_c$. No significant change of the $\nu$-values has been
  observed as long as the same $\beta_c$ was used.
\item Instead of (\ref{DISP_REL}) we have used the dispersion relation
  with $m_j^2$ replaced by $2(\cosh m_j - 1)$ (this amounts to $m_j =
  E_j$ for $p=0$). The difference between these two frequently used
  relations is on the $O(m^4)$ level and thus indicates a possible
  effect of nonleading terms.
\item We analyzed separately the masses from the $p=0$ and $p=1$
  channels.
\item We analyzed separately the masses obtained on the $16^3 32$ and
  $20^3 40$ lattices. This also corresponds to a variation of the 
  $\beta$ interval close to $\beta_c$.
\item The same data were fitted both with free $\beta_c$ and with this
  value fixed within the interval of its values obtained by other
  modifications.
\item The Gaussian and non-Gaussian group was fitted with independent
  $\beta_c$ for each group. The difference of $\beta_c$ values is
  about or less than 0.0002.
\end{enumerate}

Dropping the extremal values for each parameter, we found their values
lay in the following intervals: $\nu_{\rm ng} = 0.32$--$0.38$,
$\nu_{\rm g} = 0.41$--$0.57$ and $\beta_c = 1.1605$--$1.1609$. The
ratio $\nu_{\rm ng}/ \nu_{\rm g} = 0.62$--$0.80$ is always clearly
different from one.  These intervals are significantly broader than
the statistical uncertainties.  Assuming some peaked distribution of
the values in these intervals, we take one quarter of their widths as
an estimate of the systematic errors not accounted for by the purely
statistical analysis.

The values of the critical exponent $\nu$ given in (\ref{NUNG}),
(\ref{NUG}), (\ref{NU_RATIO}), and of $\beta_c$ given in
(\ref{BETA_C}) are the central points of these intervals. The quoted
errors are the simple sums of the above systematic errors and typical
statistical errors.

\section{Three scenarios}
The occurrence of two different correlation length exponents may seem
surprising in the light of the simulations on spherical lattices
\cite{JeLa96a,JeLa96b,LaPe96}; there the scaling of several bulk
observables could be well described by means of only one exponent,
$\nu_{\rm ng}$.  In particular, the Fisher zeros at $\gamma = -0.2$
showed no deviation from the asymptotic finite size scaling behaviour
determined by this value of the exponent $\nu$.  However, that
analysis was based completely on the bulk energy. In this observable
there is little chance of finding nonleading behaviour. The
contributions scaling with smaller critical exponents will be dominant
near the phase transition.  In terms of the eigenvalues $\lambda_i$ of
the linearized renormalization group matrix this is quite
understandable: since $\lambda_i = s^{1/\nu_i}$ (where $s$ is the
scale change factor), the smallest $\nu_i$ dominates.

Therefore we see no contradiction with the earlier results in the
appearance of two different correlation length exponents in our
present simulations.  Furthermore, it does not appear unnatural when
the possibility of the presence of a TCP at $\gamma = \gamma_0$ is
taken into account. The TCP's are known to have exponents in general
different from those associated with the adjacent ordinary critical
lines \cite{Gr73LaSa84}, and they tend to dominate the scaling
behaviour in their vicinity when the critical manifolds are approached
from outside. The cross-over regions between tricritical and critical
scaling behaviour may extend quite far away at a small angular
distance from the critical lines \cite{La76,Gr73LaSa84}.  Furthermore,
different observables, e.g.  correlation lengths, may have different
crossover regions.\footnote{We thank D.P. Landau for a helpful advice
  in this question.}

However, when trying to explain the observed scaling of the GB masses
at $\gamma = -0.2$ with help of the TCP, we see at present no reliable
possibility of saying {\em which one}, if any, of the two observed
$\nu$-values corresponds to the tricritical scaling. Instead of
guessing we formulate three scenarios, which hopefully can be tested
in future simulations. In each of them we assume that $\gamma_0 >
-0.2$.

In all the following scenarios one can consider two different
continuum limits, depending on the mass $m_j^{\rm phys} = m_j/a$ that
one chooses to fix in physical units. If it is one of the non-Gaussian
group, then the lattice constant vanishes as
\begin{equation}
         a(\tau) = \frac{m_{\rm ng}}{m_{\rm ng}^{\rm phys}}
            \propto \tau^{\nu_{\rm ng}},
\label{A_NG}
\end{equation}
and analogously for the Gaussian group,
\begin{equation}
         a(\tau) = \frac{m_{\rm g}}{m_{\rm g}^{\rm phys}}
            \propto \tau^{\nu_{\rm g}} .
\label{A_G}
\end{equation}

\vspace{0.5cm}

{\bf Scenario C:} \\ The point $\gamma = -0.2$ is a critical point
with two relevant mass scales, i.e. {\em both exponents $\nu_{\rm ng}$
  and $\nu_{\rm g}$ are critical.} Two different scaling laws at the
same critical point are unusual, but not, to our knowledge,
impossible. A somewhat analogous situation is known in the
three-dimensional pure compact U(1) gauge theory \cite{GoMa82GoMa83}.
As the ratio of masses approaches zero,
\begin{equation}
         \frac{m_{\rm g}}{m_{\rm ng}} \propto
\tau^{\nu_{\rm g} - \nu_{\rm ng}} \propto \tau^{0.5 - 0.365} \rightarrow 0,
\label{M1M2}
\end{equation}
we can consider two different continuum limits:

\begin{enumerate}
\item {\bf C$_{\rm ng}$:} Keeping $m_{\rm ng}^{\rm phys}$ constant,
  the spectrum consists of those GB states which scale with $\nu_{\rm
    ng}$ as states with finite nonzero mass, and the string tension is
  finite.  The $0^{++}$ GB would be present as a massless state.
  This would be an interesting continuum theory.
\item {\bf C$_{\rm g}$:} Keeping $m_{\rm g}^{\rm phys}$ constant, only
  the $0^{++}$ state can have a finite mass, whereas the masses of all
  other states and the string tension would run to infinity and
  decouple. Presumably this would be a Gaussian theory with a
  noninteracting scalar.
\end{enumerate}
This scenario does not make any use of the special properties of a TCP
and might hold if there is no TCP with special tricritical exponents
at $\gamma = \gamma_0$.

In the tricritical scenarios discussed next, we assume
the tricritical point at $\gamma = \gamma_0$ and take into account the
possibility that some of our scaling results are not yet asymptotic,
but obtained in a domain where a precocious form of scaling with a
``wrong'' scaling exponent is obtained.

\vspace{0.5cm} {\bf Scenario T$_{\rm g}$:} \\ The tricritical point is
Gaussian, i.e. {\em $\nu_{\rm g}$ is tricritical and $\nu_{\rm ng}$ is
  critical.} The continuum limit at $\gamma = -0.2$ is like that in
the scenario {\bf C$_{\rm ng}$} except for the mass of the $0^{++}$ GB
state. That mass shows ``wrong'' (tricritical, i.e. nonasymptotic at
this $\gamma$) scaling behaviour for those distances from $\beta_c$ we
were able to investigate on lattices of limited size. In this scenario
it would have to change its scaling behaviour closer to the phase
transition, adopting the exponent $\nu_{\rm ng}$. This state could
thus have a nonvanishing mass. The data close to $\beta_c$ suggest
that this mass would be small relatively to $m_{\rm ng}$.

At the TCP the continuum limit would be Gaussian and our present
results would say nothing about the spectrum there except the presence
of the light $0^{++}$ GB state.

\vspace{0.5cm} {\bf Scenario T$_{\rm ng}$:} \\ The tricritical point
is non-Gaussian, i.e. {\em $\nu_{\rm ng}$ is tricritical and $\nu_{\rm
    g}$ is critical.} At $\gamma = -0.2$ all the GB masses except that
of the $0^{++}$ state, as well as the string tension, show ``wrong''
(tricritical) scaling behaviour on our lattices but would change to
critical scaling when approaching the phase transition more closely.
All masses and the string tension stay finite in the continuum limit.
However, because of the Gaussian value $\nu_{\rm g}$ the theory might
be trivial with logarithmic corrections which we cannot detect.

A nontrivial continuum theory analogous to {\bf C$_{\rm ng}$} would
then be expected at the TCP. As in {\bf T$_{\rm ng}$} at $\gamma =
-0.2$ the $0^{++}$ state might have a small mass.

\vspace{1cm} At the moment it is difficult to decide which of these
scenarios is the correct one, but the most important conclusion holds
in any of them: a continuum limit is possible in which the observables
scaling like $m_{\rm ng}$ remain finite and nonzero in physical units.
Their ratios are given by the factors $r$ in front of $m_{\rm ng}$ in
eqs.~(\ref{ratios}) and (\ref{SIGMANG}), and in table
\ref{final_spectrum}. This nontrivial limit is obtained on the
critical line in the scenarios {\bf C$_{\rm ng}$} and {\bf T$_{\rm
    g}$}, whereas in the scenario {\bf T$_{\rm ng}$} it is obtained at
the hypothetical TCP. The ambiguity which remains is the mass in
physical units of the $0^{++}$ GB state in such a continuum limit: it
is zero in the {\bf C$_{\rm ng}$} scenario whereas in both the {\bf T}
scenarios it may be nonzero but small relatively to $m_{\rm ng}$.

Both scenarios {\bf T} imply that in the range of correlation lengths
we were able to investigate one of the two groups of channels shows a
behaviour at $\gamma = -0.2$ which actually corresponds to the TCP,
and not to the critical point at this $\gamma$. This can be explained
assuming a broad dominance angle of the TCP for the corresponding
states. However, we find it remarkable (peculiar?) that the would-be
critical value of $\beta$ of the group with ``false'' tricritical
behaviour agrees within our precision with that of the other group.

The special behaviour of the $0^{++}$ (strictly speaking $A_1^{++}$)
GB suggests a possible parallel with pure SU($N$) gauge theories with
a mixed action.  In that case the Wilson action, containing the trace
of the plaquette in the fundamental representation, is supplemented by
a term where the trace of the plaquette is taken in the adjoint
representation with coupling $\gamma$. The procedure is analogous to
the U(1) with extended Wilson action.  For SU(2) and SU(3) the theory
has a first order line with an endpoint at positive $\gamma$.  In the
SU(3) case it was found \cite{He95} in this region that, while the
$E^{++}$ mass remains finite, the $0^{++}$ mass goes to zero.  The
critical exponent is not known.  Other masses have not been
calculated, although it is a long-standing result for SU(2)
\cite{BhDa82} that the string tension also remains finite around the
corresponding end point.  Thus the $0^{++}$ seems to be singled out
for special behaviour as it is in the results presented here.  In
SU(3) there is presumably no phase transition for values of $\gamma$
below the endpoint, apart from a near-coincidence with finite
temperature effects on small lattices which appears to be accidental
\cite{BlDe95}\footnote{The possibility of a different interpretation,
  very similar to the phase diagram of the U(1) theory, has been
  pointed out in \cite{PaSe97}.}. It is therefore natural to suppose
the special behaviour is due to the endpoint itself. There are other
examples in lattice models where such endpoints are associated with
Gaussian behaviour (see e.g.  \cite{FrJe97b}).  Depending on the
choice from the above scenarios where the $0^{++}$ gauge-ball in the
U(1) theory scales, either the TCP or the critical line at $\gamma <
\gamma^0$ might be analogous to such endpoints.

\section{The spectrum in the Coulomb phase}

\subsection{Expected properties of photon and resonance states}
In order to prepare the ground for an understanding of our results,
let us first discuss the properties of a phase with massless photons
and massive GB's on a finite lattice. We will conclude below that this
is just what we observe from our data on the GB operator correlation
functions.

A massless state projected to zero (spatial) momentum does not decay
exponentially. From experience with Goldstone bosons
\cite{HaJa90HaJa91} we expect in fact a propagator polynomial
(parabolic) in the time variable. However, the operator projected to
non-zero momentum will have non-zero energy $E=|\vec p|\propto 1/L_s$
and an $L_s$-dependent exponential decay. This explicit finite size
dependence is an excellent indicator for massless states. Since states
with non-zero momentum are not parity-eigenstates, we expect that in
this case the photon continuum state with $J^{PC}=1^{--}$ contributes
to the $T_1^{+-}(1)$ channel. Therefore, in the $T_1^{+-}(0)$ channel
there is no signal from the photon, while the $p=1$ state has just the
energy corresponding to one unit of momentum.  Such observations thus
demonstrate the presence of a massless photon with continuum quantum
numbers $1^{--}$.

If, in addition, there is another, now massive, state e.g. in the $T_1^{+-}$
channel, it will couple to multi-photon states with these quantum
numbers and therefore will not be an asymptotic state of the system, but a
resonance. The lowest energy state for $\vec p = 0$ is the 3-photon
state with photon 3-momenta (1,0,0), (0,1,0) and (-1,-1,0) and total
energy $E=(2+\sqrt{2})$ (again in units of $2\pi/L_s$).

For simplicity of our presentation we use here the continuum
dispersion relation, although actually on the lattice (for massless
states) we have $\cosh E = 4 - \sum_{i=1}^3 \cos p_i$. In the actual
analysis and the plots we always use the lattice dispersion relation.

In a finite volume system at fixed value of the coupling we therefore
expect a specific behaviour of the (discrete) energy spectrum
\cite{Lu91aLu91b} when $L_s$ increases. For small $L_s$ the lowest
measured energy will be constant with a value close to the resonance
mass and only a weak dependence on $L_s$. At some $L_s$ the lowest
multi-photon state with energy $\propto 1/L_s$ becomes the lowest
energy state. The phenomenon of avoided level crossing will be
observed, and the lowest energy will drop inversely proportional to
the spatial extension. If one can observe higher levels in the energy
spectrum one will see the resonance as the next higher state, until
further avoided level crossings with multi-photon states of higher
momenta occur. From the energy levels one can (under certain
conditions) derive values of the phase shift in the multi-particle
channels \cite{Lu91aLu91b,GaLa93,GoKa94}

Consider now the variation with the coupling constant. The resonance
mass in lattice units then varies with the scale, whereas e.g. the
zero-momentum 3-photon energy in lattice units is still
$(2+\sqrt{2})$. The avoided level crossing therefore occurs at
a different lattice size. This behaviour provides another strong clue
for the interpretation of the spectrum.

Finally, for a given lattice size, but as a function of the coupling,
the lowest energy level in a resonance channel might increase when
moving away from the phase transition; eventually it should approach
but never exceed the lowest multi-particle energy level.  This level
decreases proportional to $1/L_s$.

\subsection{Evidence for resonances}
\begin{figure}[tbp]
  \begin{center}
    \psfig{file=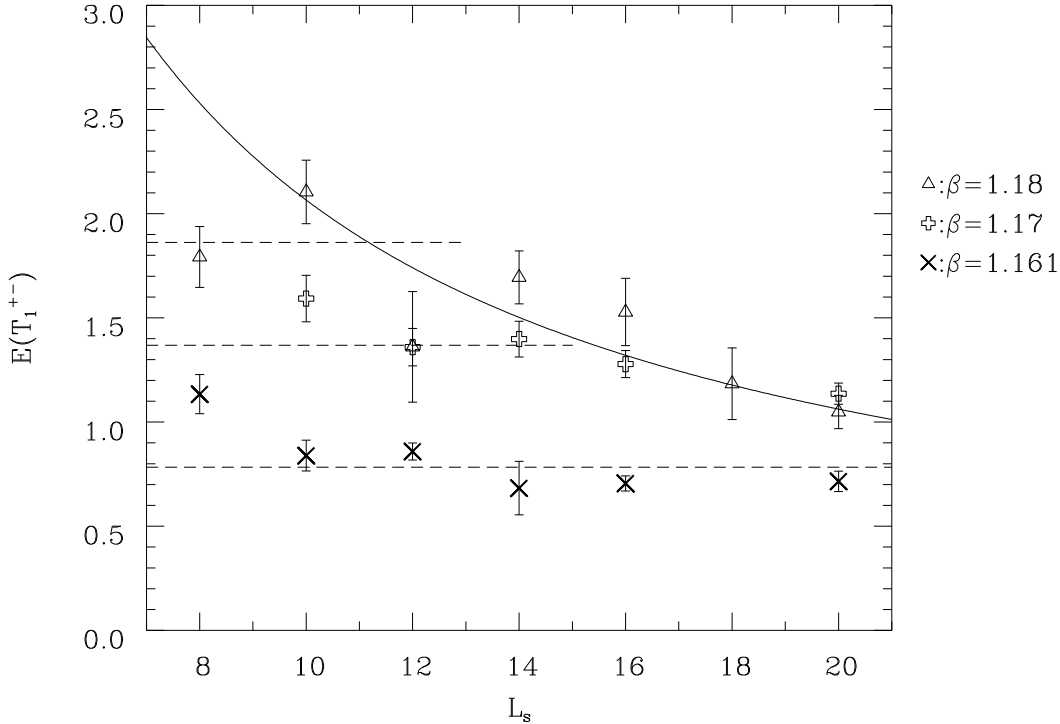,angle=90,width=\hsize,bbllx=50,
      bblly=110,bburx=545,bbury=774}
  \end{center}
  \caption{$T_1^{+-}(0)$ energy vs. 
    $L_s$ for $\beta =1.18, 1.17, 1.161$ in the Coulomb phase.  The
    full line is the energy of the 3-photon state. The dashed
    horizontal lines indicate the region of those points that have
    been included in the mean value for the determination of the
    resonance positions.}
  \label{fig:t1_l}
\end{figure}

Let us now discuss our results for gauge-balls in the Coulomb phase.
Indeed, we find a massless state in the $\vec p=(1,0,0)$ channel of
$T_1^{+-}$, with a size dependence following the expected dispersion
relation. We identify this signal with the photon. In the zero
momentum channel we find a lowest energy state following the scenario
of a resonance coupling to a 3-photon state, as discussed above. For
small lattice sizes the lowest energy level is compatible with a
constant, until it reaches the 3-photon level, which it follows for
larger spatial volumes as seen in fig.~\ref{fig:t1_l}.

We estimate the position of the resonance from the energy values
determined at lattice sizes below the point where the avoided level
crossing occurs. We indeed find a consistent behaviour indicating an
increase of the correlation length towards the phase transition.

\begin{figure}[tbp]
  \begin{center}
    \psfig{file=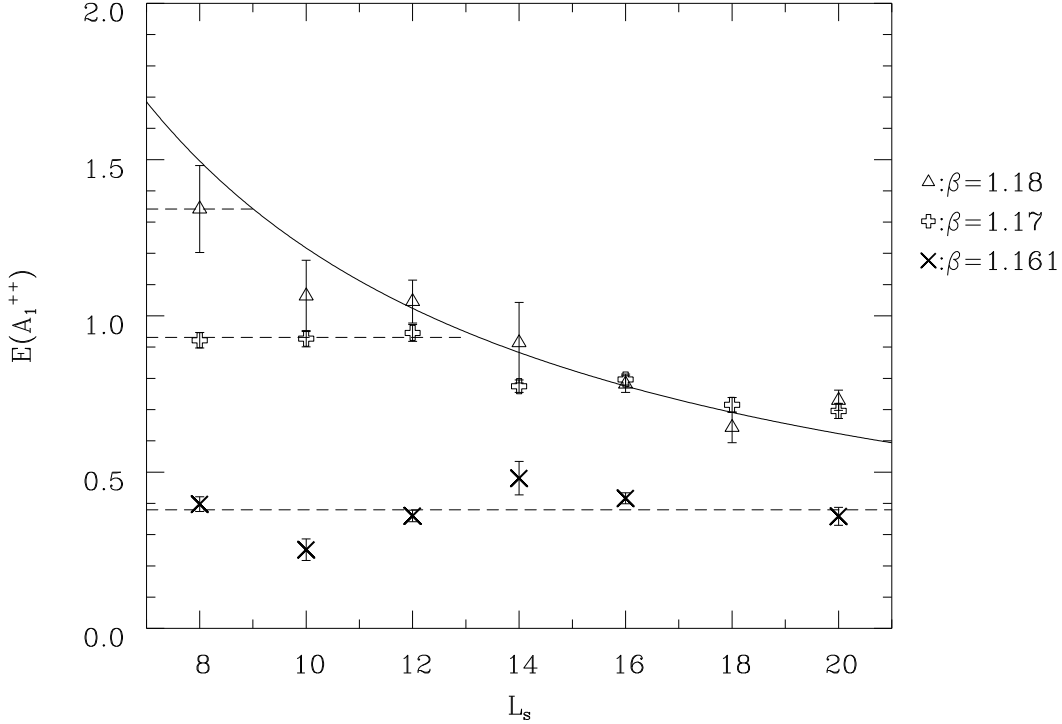,angle=90,width=\hsize,bbllx=50,
      bblly=110,bburx=545,bbury=774}
  \end{center}
  \caption{$A_1^{++}(0)$ energy vs. $L_s$ for $\beta =1.18, 1.17,
    1.161$ in the Coulomb phase. The full line is the 2-photon energy.
    The dashed horizontal lines indicate the region of those points
    that have been included in the mean value for the determination of
    the resonance positions.}
  \label{fig:a1_l}
\end{figure}

The energy levels in the channel $A_1^{++}$ exhibit a similar
behaviour (fig.~\ref{fig:a1_l}). This state is also supposedly a
resonance, since it couples to a two-photon state (photon three-momenta
(1,0,0) and (-1,0,0)) with total energy 2 (in units of $2\pi/L_s$). 

\begin{figure}[tbp]
  \begin{center}

\psfig{file=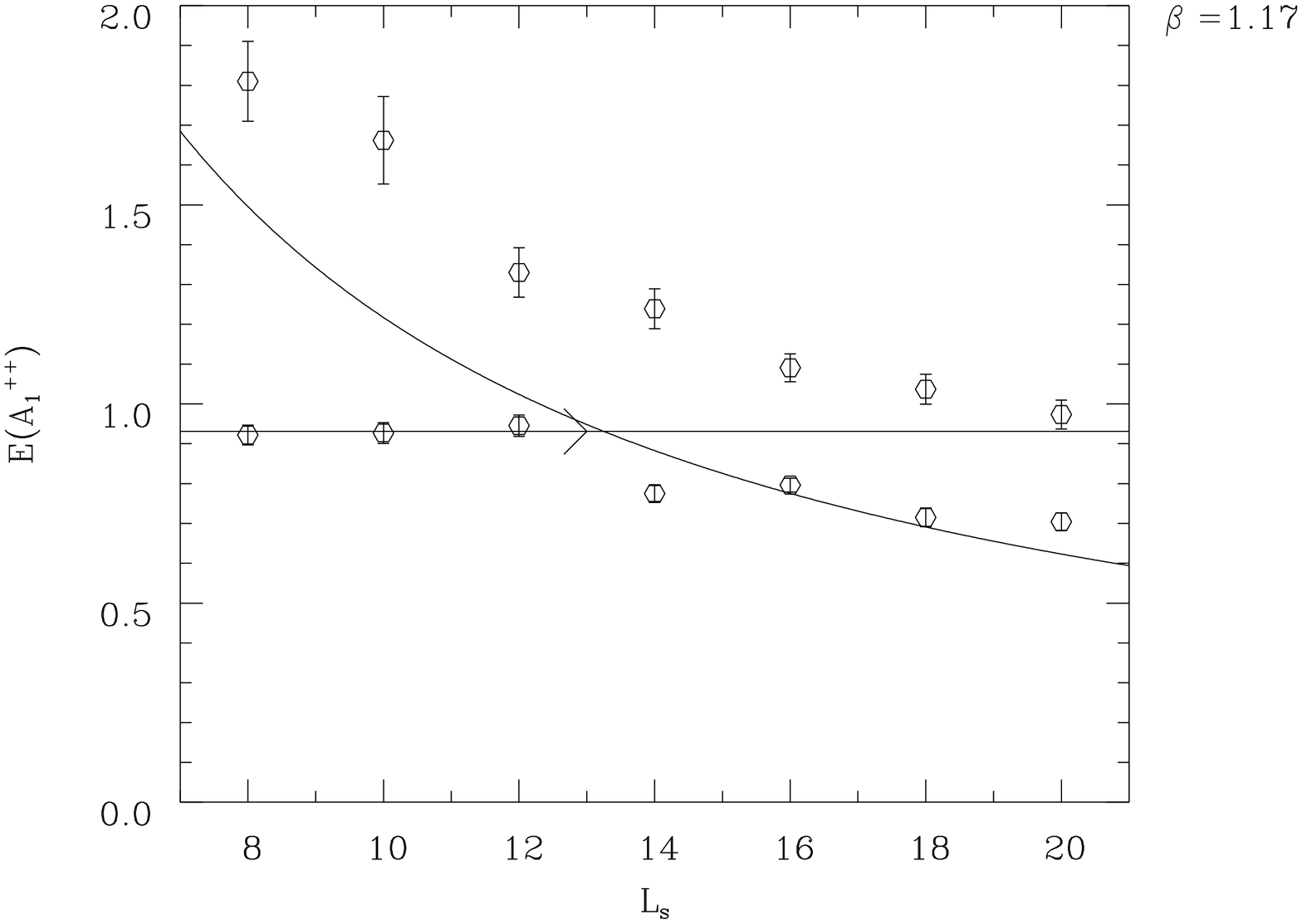,angle=90,width=\hsize,%
bbllx=50,bblly=110,bburx=545,bbury=774}
  \end{center}
  \caption{First two energy levels of the $A_1^{++}(0)$ energy  vs. $L_s$
    for beta=1.17. The horizontal line is the fit to the mass on small
    lattices. The other line represents the lowest two-photon energy.}
  \label{fig:a1_levels}
\end{figure}

In order to provide a further confirmation for the nature of this
state we have also determined the second-lowest energy level at
$\beta=1.17$ and various spatial volumes.  As discussed above, we have
usually determined the optimal operator from the diagonalization (cf.
(\ref{DIAG}) of the correlation matrix at time distance $t_0+1=1$.
Performing the diagonalization at a larger time might improve 
results for the excited states. We have tried this, but see no
good indication that the effective mass plateau for the higher states
is improved, and consequently have used $t_0=0$ again.
Fig.~\ref{fig:a1_levels}, albeit with larger errors and uncertainties
for the higher lying levels, seems to support our interpretation.

%

\begin{figure}[tbp]
  \begin{center}
    \psfig{file=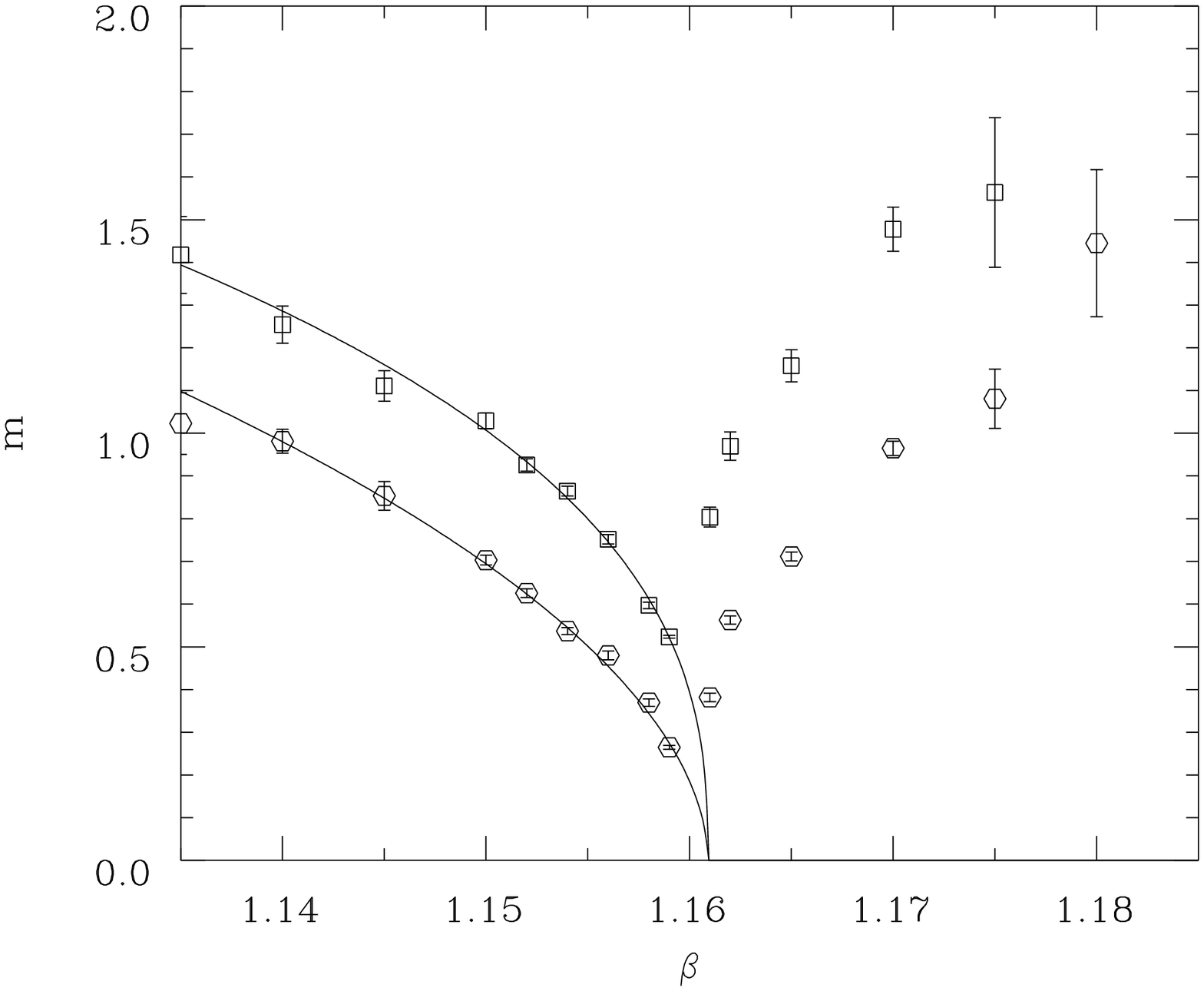,angle=90,width=\hsize,bbllx=50,
      bblly=110,bburx=545,bbury=774}
  \end{center}
  \caption{Resonance masses vs. beta in the Coulomb phase for 
    $T_1^{+-}(0)$ (squares) and $A_1^{++}(0)$ (circles). For
    comparison we show also masses in the same channels in the
    confinement phase.}
  \label{fig:scaling}
\end{figure}

Again we determine the approximate resonance position and we find
indication of critical behaviour towards the phase transition. In
fig.~\ref{fig:scaling} we give the $\beta$-dependence for the masses
of $A_1^{++}$ and $T_1^{+-}$ states.

The data is not of sufficient quality to decide on the scaling
parameters.  The determination of masses in the Coulomb phase has
various handicaps.  On one hand, according to our observations we
expect that these states are not asymptotic but resonances. Only a
narrow window is used to determine their masses. On the other hand,
\cite{BoMi93b} observed certain effects on the mass measurements of
the photon propagator due to the Dirac sheet background and related to
gauge fixing \cite{CoHe87,NaPl91}.  We do not expect these to affect
our results noticeably since we have comparatively large spatial
lattices.  Finally, due to the massless state the finite size effects
are definitely larger than in the confinement phase. We think that
this last problem may be the main reason for our difficultiess to
identify a consistent scaling form from the Coulomb side.

\begin{figure}[tbp]
  \begin{center}
    \psfig{file=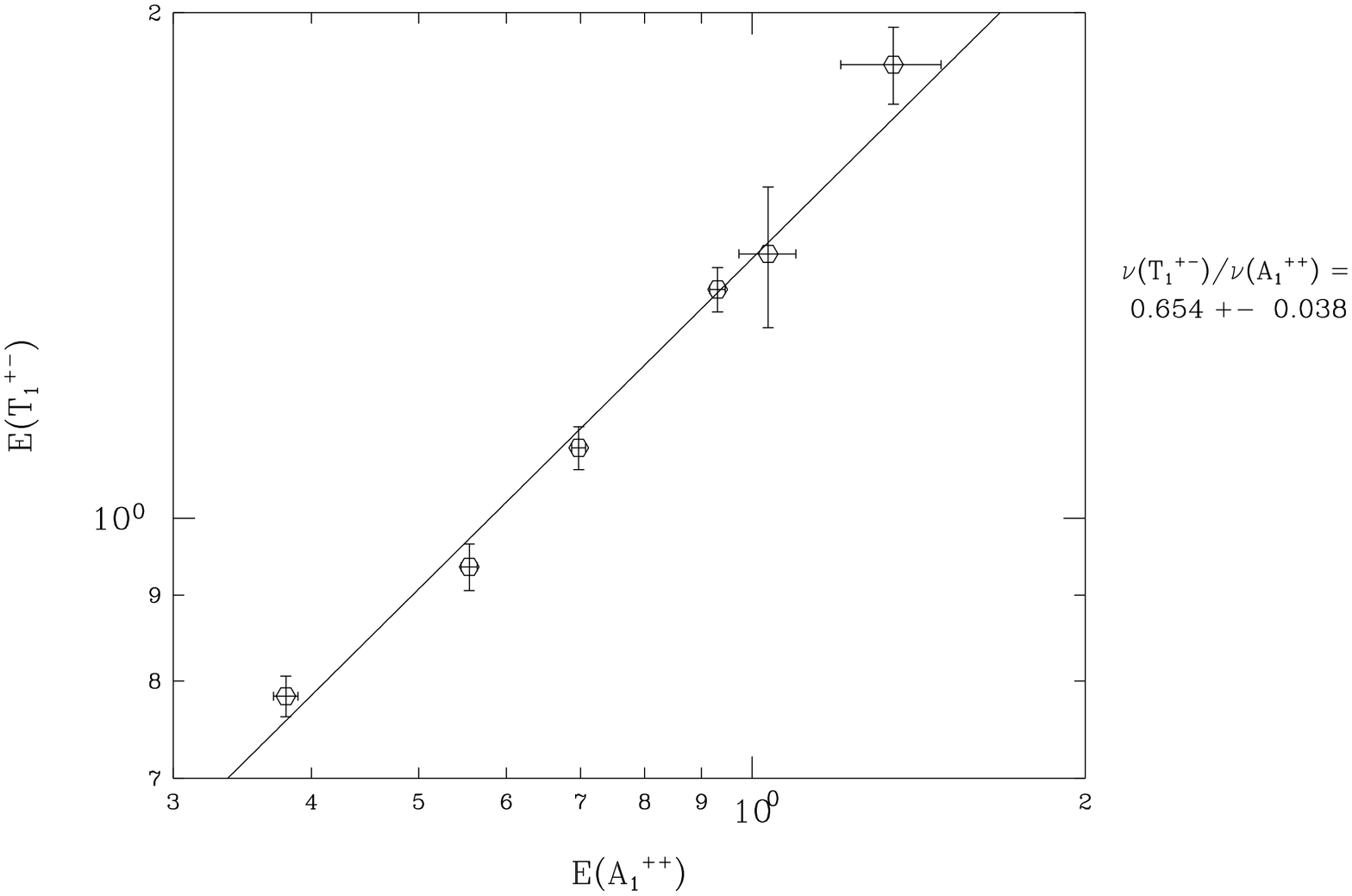,angle=90,width=\hsize,bbllx=50,
      bblly=110,bburx=545,bbury=774}
  \end{center}
  \caption{Different scaling of the two gauge-ball states observed 
    in the Coulomb phase, assuming that the systematic uncertainties
    are less important in the ratio of the two masses.}
  \label{fig:log_log_coulomb}
\end{figure}

Our results indicate that in the Coulomb phase one has both massless
vector states (photons) and massive GB resonances $A_1^{++}$ and
$T_1^{+-}$ that couple to 2 or 3 of the massless states, respectively.
The masses of the resonance states seem to scale towards the phase
transition. We cannot decide whether both scale differently or not.
However, assuming that the uncertainties associated with the
determination of each individual mass partly cancel in their ratio,
one can look at whether the log-log plot, analogous to
fig.~\ref{fig:loglog} in the confinement phase, has a slope different
from one. In fig.~\ref{fig:log_log_coulomb} we see that the data
indeed indicate the slope being consistent with that in
fig.~\ref{fig:loglog}. This could point towards the existence of two
mass scales in the Coulomb phase, too. 

As for the other channels, our data suggest that the
$T_2^{+-}(0), A_2^{+-}(0)$ and $E^{+-}(0)$ behave similar to
$T_1^{+-}(0)$. Thus massive three-photon resonances, possibly with the
same or similar mass, could be present in all these channels. The
spectrum of the resonances in the Coulomb phase might thus resemble
that of the gauge-balls in the confinement phase.

\section{Conclusions and open questions}

Our results in the confinement phase are in agreement with the earlier
finite size studies on spherical lattices
\cite{JeLa96a,JeLa96b,LaPe96} but substantially extend that work.
They strongly suggest the existence of a continuum quantum field
theory in four dimensions with the following properties:

\begin{enumerate}
\item From the confining phase one approaches a non-Gaussian fixed
  point, so the theory is presumably interacting but not
  asymptotically free. The correlation length exponent at this fixed
  point is $\nu_{\rm ng} \simeq 0.36$. This value is obtained by
  combining our present results with those in
  \cite{JeLa96a,JeLa96b,LaPe96}.
\item The well measurable physical states at this fixed point
  contribute to the $A_2, E, T_1$ and $T_2$ representations of the
  lattice symmetry group with $PC = +-$. They have equal or very
  similar mass $m_{\rm ng}^{\rm phys}$. Assuming that the smallest
  spin dominates, the continuum quantum numbers of gauge-balls would
  be $R^{PC} = 1^{+-}, 2^{+-}, 3^{+-}$. This assignment is uncertain
  as it could be that for example the $3^{+-}$ GB dominates the
  $T_1^{+-}$ channel and the $1^{+-}$ state is not present.
\item Depending on the scenario for the continuum limit of the lattice
  theory, the continuum theory might contain a light (compared to
  $m_{\rm ng}^{\rm phys}$), possibly massless $A_1^{++} = 0^{++}$
  gauge-ball. However, it is probably only weakly coupled to the
  states in other channels, and might be completely decoupled. No
  other state in the $A_1^{++}$ channel was observed.
\item The theory is confining in the sense of the Wilson criterium,
  i.e. the string tension in physical units $\sigma^{\rm phys}$ has a
  finite nonvanishing value $\sqrt{ \sigma^{\rm phys}} \simeq 0.34
  m_{\rm ng}^{\rm phys} $.
\item In the $E^{++}$ and $T_2^{++}$ (presumably $2^{++}$) channels a
  state with $m_{2^{++}}^{\rm phys} \simeq 2m_{\rm ng}^{\rm phys}$ is
  present. It may be a two-particle state.
\item In many other channels observable states with masses $\simeq
  (2.6 - 3.6)m_{\rm ng}^{\rm phys}$ are present. Some or all of them
  might be two- or three-particle states.
\item No effective energy plateau was found in the channels $A_1^{+-},
  A_2^{-+}, E^{-+}$ and $T_1^{-+}$. However, even in these channels
  the effective energy at the distance $t=0$ is measurable, indicating
  the possible presence of some heavy, possibly multiparticle states.
\end{enumerate}

We point out that because of the rigorous duality relations for the
U(1) lattice gauge theory the same continuum theory can also be
obtained as the continuum limit of the following four-dimensional
lattice theories: Coulomb gas of monopole loops \cite{BaMy77}, $\mathbb{Z}$
gauge theory \cite{FrMa86FrMa87}, and non-compact U(1) Higgs model at
large negative squared bare scalar mass (frozen four-dimensional
superconductor) \cite{Pe78,FrMa86FrMa87}. The last model can be
understood as a limit of a theory described by world sheets of
Nielsen-Olesen strings \cite{PoSt91PoWi93ArBa94}.

Currently we have no physical picture explaining the observed
gauge-ball spectrum. It could be that its understanding might rather
come from one of the dual equivalents, for example as the lowest
states of a closed Nielsen-Olesen string.

Our results in the Coulomb phase indicate that also here a continuum
theory approaching the confinement transition might exist which
differs from the free Maxwell theory usually expected in this phase at
least in the $\beta\to\infty$ limit.  However, the masses and the
scaling behaviour of the resonances observed in this phase are not yet
fully understood. One or even two mass scales might emerge there.

As a spin-off of our study of this phase we confirm the
universality \cite{Ca80,Lu82,JeNe85} of the maximal value of the
renormalized coupling by obtaining $\alpha_{\rm R}^{\rm max} =
0.19(1)$ in agreement with results in other lattice formulations of
the U(1) gauge theory.

Our present work leaves unanswered several questions about the pure
U(1) lattice gauge theory and raises many new ones. The major challenge
is to understand how the order of the confinement-Coulomb phase
transition changes from first at $\gamma > \gamma_0$ to second at
$\gamma < \gamma_0$. Is the point $\gamma_0$ a tricritical point in
the sense that several critical lines emerge from it? What are the
additional couplings required to reveal this structure? Where are the
crossover regions between critical and tricritical scaling behaviour?
Are these regions for the $0^{++}$ state different from those of the
other GB states, as required by the {\bf T$_{\rm ng}$} and {\bf
  T$_{\rm g}$} scenarios? Or is $\gamma_0$ an ordinary point of the
critical line $\gamma \le \gamma_0$, as would be most natural in the
scenario {\bf C}? Is $\nu_{\rm ng}$ a critical or tricritical
exponent?  A clarification of these questions will require large
effort, as $\gamma$ and possibly some other couplings will have to be
varied.

Some arguments can be given for prefering the scenario {\bf T$_{\rm
    ng}$}: (i) It is consistent with some MCRG studies
\cite{La86La87bLaRe87}.  (ii) Previous results for the string tension
\cite{JeNe85} and for the massive photon mass \cite{CoHe87} at $\gamma
= 0$, i.e. very close to $\gamma_0$, are consistent with the
non-Gaussian value. (iii) It is a fairly conventional scenario
implying a broad range of dominance of the TCP.

However, the observation of $\nu \simeq 0.36$ in the finite size
scaling analysis even at $\gamma = -0.5$ \cite{JeLa96a,JeLa96b} could
be difficult to accommodate in that scenario and prefers scenario {\bf
  C}. Two different scaling laws at the same critical point might seem
to be somewhat exotic. However, the rigorously established results in
the three-dimensional pure U(1) gauge theory \cite{GoMa82GoMa83} are
strikingly similar. Also there a $0^{++}$ mass scales faster than the
string tension when the gauge coupling vanishes (both scale
exponentially).  This implies the existence of two different continuum
limits. The one with confinement contains a massless scalar, too.

We stress that the existence of the non-Gaussian continuum theory does
not depend on which scenario is correct. Only the properties of the
$0^{++}$ state, whether it is light but massive or massless, and
whether it couples or not to the other states, depend on the scenario.
Thus the fate of the $0^{++}$ state in the non-Gaussian continuum
limit is a challenging question.

The particular scaling behaviour of the $0^{++}$ GB observed in
the present work, in the lattice QCD \cite{He95} and in the
three-dimensional pure U(1) lattice gauge theory \cite{GoMa82GoMa83}
should draw attention towards light scalars in the confining theories.

In the light of so many open questions we should mention the natural
caveat of numerical simulations: it might be that only much closer to
the phase transition, on lattices much larger than we were able to
use, the true behaviour will show up in the future. For example, the
scaling of all observables might turn out to be the same at any
$\gamma \le \gamma_0$, or the first order transition might reappear
even on spherical lattices.

There are several paths which could be pursued with the means
currently available:
\begin{itemize}
\item Studies similar to the present one at various $\gamma$ could
  elucidate the fate of the $0^{++}$ state and show the cross-over
  regions if they exist. For this purpose both large negative and some
  positive $\gamma$ values will be required.
\item It should be possible to investigate the short-range form of the
  static potential in the confinement phase. This would shed more
  light on the nontrivial field theory strongly interacting at short
  distances.
\item It would be interesting to find out whether the heavier states
  observed in the confinement phase are genuine gauge-balls or
  multi-particle states. If they are resonances, they should be
  investigated with the appropriate finite size techniques
  \cite{Lu91aLu91b,GaLa93,GoKa94}.
\item An analysis of the resonances in the Coulomb phase with these
  techniques would establish their existence and allow to determine
  the scaling behaviour and continuum limit in this phase as well. The
  obstacle is the presence of the massless photon, which complicates
  the analysis devised for theories with a mass gap
  \cite{Lu91aLu91b}.
\item Simulations of the dual $\mathbb{Z}$ theory might bring further
  insight into the finite size properties, scaling behaviour of the
  monopole mass \cite{PoWi91}, role of boundary conditions, etc.
\end{itemize}
The pure gauge theory investigated here may not be realized in nature.
However, it seems worthwhile to pursue its study, since it widens our
understanding of quantum field theories.  Our lattice simulations have
brought about unexpected results. Further surprises may be waiting.

\begin{ack}
  We thank V.~Dohm, M.~G\"ockeler, S.~Hands, H.A.~Kastrup,
  D.P.~Landau, M.~L\"uscher, G.~Mack, S.~Meyer, G.~M\"unster,
  E.~Seiler, P.~Weisz, and J.~Zinn-Justin for discussions. The
  computations have been performed on the Cray-YMP and Cray-T90 of
  HLRZ J\"ulich. A supplementary Cray-T90 time grant is gratefully
  acknowledged. J.C., W.F. and J.J. thank HLRZ J\"ulich for
  hospitality. The work was supported by DFG.
\end{ack}

\bibliographystyle{wunsnot}   

\vfill
\pagebreak
\appendix
\section{Appendix}

In this appendix we present two tables with more detailed information
about the GB masses. Further data on the effective energies in the
confinement phase can be obtained from the authors by e-mail
(jersak@physik.rwth-aachen.de).

\begin{table}[htb]
  \caption
  {The effective energies of the $A_1^{++}$ and $T_1^{+-}$ gauge-balls
    with zero momentum in the confinement phase on the time distances
    $t/(t+1)$.}
  \label{tab:effmass}
  \vspace{0.5cm}
  \scriptsize

    \begin{center}
    \leavevmode
    \scriptsize
    \begin{tabular}{|c|c|c|c|c|c|c|c|c|}\hline
      \multicolumn{9}{|c|}{$A_1^{++}$}\\
      \hline\hline
     Gitter & $\beta$ & $0/1$ & $1/2$ & $2/3$ & 3/4&4/5&4/6&6/7\\
      \hline  \hline
&         1.13 &    1.159(8)&     1.09(2)&     1.05(6)&     0.90(16)&      1.1(7)&      1.3(11)&  \\ 
      \cline{2-9}  
&        1.135 &    1.074(8)&    1.023(20)&     0.96(7)&     1.05(15)&      1.2(8)&      1.4(11)&  \\ 
      \cline{2-9}  
&         1.14 &    0.968(6)&    0.914(14)&     0.92(3)&     1.05(11)&      0.7(3)&      0.4(3)&      0.2(5) \\ 
      \cline{2-9}  
&        1.145 &    0.862(6)&    0.815(14)&     0.84(4)&     0.88(8)&     0.76(16)&      0.7(3)&      0.1(2) \\ 
      \cline{2-9}  
      $16^332$
&         1.15 &    0.732(4)&    0.692(9)&    0.701(15)&     0.73(4)&     0.77(9)&      0.7(2)&      0.4(4) \\ 
      \cline{2-9}  
&        1.152 &    0.654(6)&    0.618(10)&    0.595(15)&     0.58(3)&     0.60(5)&     0.60(11)&     0.41(13) \\ 
      \cline{2-9}  
&        1.154 &    0.607(8)&    0.559(12)&    0.557(18)&     0.53(3)&     0.50(5)&     0.47(9)&     0.42(15) \\ 
      \cline{2-9}  
&        1.156 &    0.511(7)&    0.472(10)&    0.468(15)&     0.46(2)&     0.48(3)&     0.46(5)&     0.50(8) \\ 
      \cline{2-9}
&        1.158 &    0.412(10)&    0.375(12)&    0.376(16)&     0.38(2)&     0.39(3)&     0.45(5)&     0.48(8) \\ 
      \hline  \hline
&        1.154 &    0.588(10)&    0.544(13)&    0.530(18)&     0.53(3)&     0.57(6)&     0.68(17)&      1.1(8) \\ 
      \cline{2-9}  
&        1.156 &    0.533(11)&    0.490(17)&     0.50(3)&     0.51(5)&     0.54(9)&     0.54(19)&      0.5(3) \\ 
      \cline{2-9}  
      \raisebox{1.5ex}[-1.5ex]{$20^340$}
&        1.158 &    0.410(10)&    0.365(14)&    0.355(19)&     0.35(3)&     0.33(4)&     0.30(5)&     0.28(6) \\ 
      \cline{2-9}  
&        1.159 &    0.306(5)&    0.271(6)&    0.268(7)&    0.262(8)&    0.263(10)&    0.262(12)&    0.260(16) \\ 
      \cline{2-9}  
       \hline
    \end{tabular}
  \end{center}

  \vspace{0.4cm}

    \begin{center}
    \leavevmode
    \scriptsize
    \begin{tabular}{|c|c|c|c|c|c|c|c|c|}\hline
      \multicolumn{9}{|c|}{$T_1^{+-}$}\\
      \hline\hline
     Gitter & $\beta$ & $0/1$ & $1/2$ & $2/3$ & 3/4&4/5&4/6&6/7\\
      \hline  \hline
&         1.13 &    1.496(6)&     1.43(2)&     1.32(12)&      1.5(4)& & &  \\ 
      \cline{2-9}  
&        1.135 &    1.384(5)&    1.314(20)&     1.33(8)&      1.1(3)&      1.4(11)& &  \\ 
      \cline{2-9}  
&         1.14 &    1.293(5)&    1.236(16)&     1.24(6)&     1.17(17)&      0.6(4)&      0.3(5)&      1.4(9) \\ 
      \cline{2-9}  
&        1.145 &    1.189(4)&    1.121(12)&     1.08(3)&     0.99(10)&      0.9(2)&      0.4(4)&      0.4(5) \\ 
      \cline{2-9}  
      $16^332$
&         1.15 &    1.066(4)&    0.981(9)&     0.97(3)&     0.92(6)&     0.80(15)&      0.6(3)&  \\ 
      \cline{2-9}  
&        1.152 &    1.006(4)&    0.925(8)&    0.903(19)&     0.90(6)&     0.94(16)&      1.1(9)&  \\ 
      \cline{2-9}  
&        1.154 &    0.950(4)&    0.863(8)&    0.843(18)&     0.87(4)&     0.87(12)&      1.2(5)&      0.4(7) \\ 
      \cline{2-9}  
&        1.156 &    0.798(4)&    0.738(8)&    0.732(14)&     0.75(3)&     0.73(5)&     0.71(14)&      1.0(5) \\ 
      \cline{2-9}
&        1.158 &    0.668(4)&    0.601(6)&    0.588(10)&    0.564(17)&     0.56(3)&     0.57(6)&     0.54(10) \\ 
      \hline  \hline
&        1.154 &    0.956(5)&    0.862(13)&     0.83(3)&     0.76(5)&     0.73(15)&      0.5(2)&      0.6(6) \\ 
      \cline{2-9}  
&        1.156 &    0.779(6)&    0.729(11)&     0.74(3)&     0.73(5)&     0.81(11)&      0.7(2)&      0.8(5) \\ 
      \cline{2-9}  
      \raisebox{1.5ex}[-1.5ex]{$20^340$}
&        1.158 &    0.676(5)&    0.624(9)&    0.624(18)&     0.60(3)&     0.60(5)&     0.58(11)&      0.7(2) \\ 
      \cline{2-9}  
&        1.159 &   0.5933(19)&    0.528(3)&    0.518(4)&    0.510(6)&    0.528(11)&    0.540(16)&     0.53(3) \\ 
      \cline{2-9}  
       \hline
    \end{tabular}
  \end{center}
\end{table}


\begin{table}[ht]
  \caption
  {Masses for all $R^{PC}$ combinations obtained from operators with zero and 
    smallest nonvanishing momentum.}
  \label{tab:mass1}
  \vspace{0.5cm}
  \begin{center}
    \leavevmode
    \scriptsize
    \begin{tabular}{|c|c|c|c|c|c|c|c|c|}\hline
      & &\multicolumn{7}{|c|}{$16^332$}\\
      \cline{3-9}
      
      \raisebox{1.5ex}[-1.5ex]{$R^{PC}$}& \raisebox{1.5ex}[-1.5ex]{$\frac{p_zL}{2\pi}$} & $\beta=1.13$& $\beta=1.135$& $\beta=1.14$& $\beta=1.145$& $\beta=1.15$& $\beta=1.152$& $\beta=1.154$ \\
      \hline  \hline
      &                                    0&    1.08(7)  &    1.02(8)  &    0.95(4)  &    0.85(5)  &   0.721(20) &   0.599(19) &    0.55(2)   \\ 
      \cline{2-9}                                                                                                                                         
      \raisebox{1.5ex}[-1.5ex]{$A_1^{++}$}&1&    1.27(11) &    1.25(8)  &    1.00(4)  &    0.80(3)  &    0.70(3)  &    0.66(2)  &   0.566(20)  \\ 
      \hline                                                                                                                                  
      $A_1^{+-}$&                          0&             &             &     2.3(12) &     3.9(18) &     2.1(10) &     1.8(4)  &     2.1(6)   \\ 
      \hline                                                                                                                                  
      $A_1^{-+}$&                          0&     2.6(17) &             &             &             &     2.4(6)  &     2.7(17) &     2.7(8)   \\ 
      \hline                                                                                                                                  
      $A_1^{--}$&                          0&       3(2)  &             &     1.0(11) &     1.4(12) &     2.1(7)  &     1.8(14) &     2.1(7)   \\ 
      \hline                                                                                                                                  
      $A_2^{++}$&                          0&     2.2(14) &     2.5(16) &             &       5(4)  &     3.5(19) &       5(3)  &     1.9(4)   \\ 
      \hline                                                                                                                                  
      $A_2^{+-}$&                          0&     1.5(3)  &     1.6(3)  &    1.20(10) &    1.19(9)  &    1.00(4)  &    0.95(5)  &    0.92(4)   \\ 
      \hline                                                                                                                                  
      $A_2^{-+}$&                          0&     1.3(13) &             &             &     1.9(11) &       3(2)  &       5(3)  &              \\ 
      \hline                                                                                                                                  
      $A_2^{--}$&                          0&             &     3.2(20) &             &      11(8)  &     3.6(17) &             &     1.2(6)   \\ 
      \hline                                                                                                                                  
      &                                    0&     2.2(5)  &     1.8(9)  &     3.0(5)  &     2.6(3)  &    2.10(14) &    1.88(12) &    1.74(8)   \\ 
      \cline{2-9}                                                                                                                             
      \raisebox{1.5ex}[-1.5ex]{$E^{++}$}&  1&     1.4(14) &     2.8(7)  &     2.5(3)  &     2.6(3)  &    2.25(17) &    2.04(13) &    1.99(11)  \\ 
      \hline                                                                                                                                  
      $E^{+-}$&                            0&    1.65(19) &    1.29(12) &    1.31(11) &    1.05(6)  &    0.96(4)  &    0.91(3)  &    0.84(3)   \\ 
      \hline                                                                                                                                  
      $E^{-+}$&                            0&             &       6(2)  &     1.2(16) &     1.9(9)  &     4.0(16) &     2.9(6)  &     3.0(7)   \\ 
      \hline                                                                                                                                  
      $E^{--}$&                            0&      10(4)  &       7(5)  &             &       5(3)  &     2.2(6)  &     2.3(6)  &     2.7(6)   \\ 
      \hline                                                                                                                                  
      &                                    0&             &       4(2)  &      10(5)  &             &       4(2)  &       9(4)  &     1.4(9)   \\ 
      \cline{2-9}                                                                                                                             
      \raisebox{1.5ex}[-1.5ex]{$T_1^{++}$}&1&             &       5(2)  &             &             &     2.3(12) &             &     1.4(5)   \\ 
      \hline                                                                                                                                  
      &                                    0&    1.43(15) &    1.42(9)  &    1.29(7)  &    1.11(4)  &    1.00(3)  &    0.93(2)  &    0.87(2)   \\ 
      \cline{2-9}                                                                                                                             
      \raisebox{1.5ex}[-1.5ex]{$T_1^{+-}$}&1&    1.58(19) &    1.25(16) &    1.50(12) &    1.16(9)  &    1.01(4)  &    0.85(4)  &    0.88(4)   \\ 
      \hline                                                                                                                                  
      &                                    0&     2.4(15) &     2.5(15) &     1.4(8)  &       5(3)  &      13(5)  &     1.8(9)  &       4(2)   \\ 
      \cline{2-9}                                                                                                                             
      \raisebox{1.5ex}[-1.5ex]{$T_1^{-+}$}&1&     1.7(4)  &     1.0(3)  &    1.13(11) &    0.76(7)  &    0.73(3)  &    0.67(4)  &    0.59(3)   \\ 
      \hline                                                                                                                                  
      &                                    0&     2.6(17) &       4(2)  &      30(13) &     1.6(8)  &     2.7(11) &     1.7(7)  &     3.9(15)  \\ 
      \cline{2-9}                                                                                                                             
      \raisebox{1.5ex}[-1.5ex]{$T_1^{--}$}&1&     0.8(14) &     1.4(6)  &     1.6(11) &             &     2.7(14) &       4(2)  &     1.7(14)  \\ 
      \hline                                                                                                                                  
      &                                    0&     3.6(11) &     3.3(6)  &     3.1(3)  &     2.6(2)  &    2.13(11) &    1.91(8)  &    1.85(6)   \\ 
      \cline{2-9}                                                                                                                             
      \raisebox{1.5ex}[-1.5ex]{$T_2^{++}$}&1&     4.8(20) &     2.5(4)  &     4.4(16) &     2.3(2)  &     2.2(2)  &    1.95(14) &    1.73(8)   \\ 
      \hline                                                                                                                                  
      &                                    0&    1.48(14) &    1.44(10) &    1.25(8)  &    1.07(5)  &    0.99(3)  &    0.93(2)  &    0.86(2)   \\ 
      \cline{2-9}                                                                                                                             
      \raisebox{1.5ex}[-1.5ex]{$T_2^{+-}$}&1&    1.37(19) &    1.14(14) &    1.45(14) &    1.28(9)  &    0.99(5)  &    0.92(5)  &    0.83(4)   \\ 
      \hline                                                                                                                                  
      &                                    0&       6(3)  &     2.6(12) &     2.3(10) &     1.7(14) &     3.9(12) &     2.4(4)  &     2.1(2)   \\ 
      \cline{2-9}                                                                                                                             
      \raisebox{1.5ex}[-1.5ex]{$T_2^{-+}$}&1&             &             &     1.7(4)  &     1.8(13) &     3.0(9)  &     2.4(6)  &     2.4(4)   \\ 
      \hline                                                                                                                                  
      &                                    0&             &             &       7(3)  &      19(10) &       7(4)  &     2.8(17) &     2.8(7)   \\ 
      \cline{2-9}                                                                                                                             
      \raisebox{1.5ex}[-1.5ex]{$T_2^{--}$}&1&     2.9(19) &       3(2)  &     1.0(4)  &     1.4(7)  &     0.9(2)  &    1.09(15) &    0.86(9)   \\ 
      \hline                                  
    \end{tabular}                            
  \end{center}                               

\end{table}


\begin{table}[ht]
  \caption
  {Continuation of table \ref{tab:mass1}}
  \label{mass2}
  \vspace{0.5cm}

  \begin{center}                             
    \leavevmode                              
    \scriptsize                            
    \begin{tabular}{|c|c|c|c||c|c|c|c|}
      \hline
      & &\multicolumn{2}{|c||}{$16^332$}&\multicolumn{4}{|c|}{$20^340$}\\
      \cline{3-8}                            
                                             
      \raisebox{1.5ex}[-1.5ex]{$R^{PC}$}&    
      \raisebox{1.5ex}[-1.5ex]{$\frac{p_zL}{2 \pi}$} & $\beta=1.156$&
      $\beta=1.158$& $\beta=1.154$& $\beta=1.156$& $\beta=1.158$&
      $\beta=1.159$  \\        
      \hline  \hline                         
      &                                    0&    0.47(2)  &    0.39(2)  &    0.54(2)  &    0.51(4)  &    0.34(3)  &   0.265(9)    \\ 
      \cline{2-8}                                                                                                                             
      \raisebox{1.5ex}[-1.5ex]{$A_1^{++}$}&1&   0.459(19) &    0.33(3)  &    0.51(2)  &    0.43(3)  &    0.31(2)  &   0.293(6)    \\  
      \hline                                                                                                                                  
      $A_1^{+-}$&                          0&    1.88(18) &    1.70(11) &     2.3(7)  &     1.8(9)  &     1.6(2)  &    1.65(4)   \\ 
      \hline                                                                                                                                  
      $A_1^{-+}$&                          0&     2.7(6)  &    1.72(14) &      11(6)  &     1.9(6)  &     2.0(3)  &    1.49(3)   \\ 
      \hline                                                                                                                                  
      $A_1^{--}$&                          0&     2.6(7)  &    1.73(20) &     1.6(14) &     1.2(3)  &     1.3(4)  &    1.47(4)   \\ 
      \hline                                                                                                                                  
      $A_2^{++}$&                          0&     2.3(13) &     2.0(2)  &       4(2)  &     2.2(8)  &     2.1(3)  &    1.70(5)   \\ 
      \hline                                                                                                                                              
      $A_2^{+-}$&                          0&    0.79(3)  &    0.61(2)  &    0.84(6)  &    0.71(4)  &    0.62(3)  &   0.529(10)  \\ 
      \hline                                                                                                                                  
      $A_2^{-+}$&                          0&             &     1.5(4)  &     1.0(17) &     0.8(9)  &             &     2.8(4)   \\ 
      \hline                                                                                                                                  
      $A_2^{--}$&                          0&     2.4(9)  &     1.8(2)  &     1.9(12) &     2.1(10) &     1.8(4)  &    1.46(5)   \\ 
      \hline                                                                                                                                  
      &                                    0&    1.56(5)  &    1.22(2)  &    1.70(9)  &    1.60(8)  &    1.30(5)  &   1.098(11)  \\ 
      \cline{2-8}                                                                                                                             
      \raisebox{1.5ex}[-1.5ex]{$E^{++}$}&  1&    1.67(7)  &    1.47(4)  &    1.90(12) &    1.80(13) &    1.42(6)  &   1.193(12)  \\ 
      \hline                                                                                                                                  
      $E^{+-}$&                            0&    0.76(2)  &   0.590(17) &    0.85(4)  &    0.76(4)  &    0.62(3)  &   0.510(8)   \\ 
      \hline                                                                                                                                  
      $E^{-+}$&                            0&     2.1(3)  &    1.63(10) &     1.9(13) &     1.5(2)  &    1.66(15) &    1.58(3)   \\ 
      \hline                                                                                                                                  
      $E^{--}$&                            0&    2.02(19) &    1.78(10) &     1.6(4)  &     2.1(4)  &    1.77(15) &    1.56(3)   \\ 
      \hline                                                                                                                                  
      &                                    0&     0.9(7)  &     2.4(3)  &     2.8(15) &      10(5)  &     1.8(14) &    1.94(6)   \\ 
      \cline{2-8}                                                                                                                              
      \raisebox{1.5ex}[-1.5ex]{$T_1^{++}$}&1&     3.0(11) &     1.9(3)  &       6(3)  &     1.4(8)  &     2.9(12) &    1.89(9)   \\ 
      \hline                                                                                                                                  
      &                                    0&   0.751(16) &   0.588(12) &    0.83(3)  &    0.76(4)  &    0.63(2)  &   0.524(5)   \\ 
      \cline{2-8}                                                                                                                             
      \raisebox{1.5ex}[-1.5ex]{$T_1^{+-}$}&1&    0.76(3)  &    0.64(2)  &    0.78(5)  &    0.71(3)  &    0.62(2)  &   0.522(6)   \\ 
      \hline                                                                                                                                  
      &                                    0&     1.8(2)  &     1.9(4)  &     2.5(14) &             &     1.6(3)  &    2.22(9)   \\ 
      \cline{2-8}                                                                                                                             
      \raisebox{1.5ex}[-1.5ex]{$T_1^{-+}$}&1&    0.45(2)  &    0.33(3)  &    0.51(3)  &    0.45(4)  &    0.33(2)  &   0.295(8)   \\ 
      \hline                                                                                                                                  
      &                                    0&    1.81(17) &    1.62(9)  &     3.0(15) &     1.9(4)  &    1.82(16) &    1.50(3)   \\ 
      \cline{2-8}                                                                                                                             
      \raisebox{1.5ex}[-1.5ex]{$T_1^{--}$}&1&     1.7(2)  &
      2.09(18) &     2.3(8)  &     1.6(8)  &     1.7(2)  &    1.59(3)
      \\ 
      \hline                                                                                                                                  
      &                                    0&    1.68(5)  &    1.29(3)  &    1.75(9)  &    1.53(6)  &    1.26(4)  &   1.077(8)   \\ 
      \cline{2-8}                                                                                                                             
      \raisebox{1.5ex}[-1.5ex]{$T_2^{++}$}&1&    1.64(6)  &    1.34(4)  &    1.66(12) &    1.55(10) &    1.31(5)  &   1.209(12)  \\ 
      \hline                                                                                                                                  
      &                                    0&   0.761(16) &   0.591(12) &    0.85(3)  &    0.76(4)  &    0.63(2)  &   0.525(7)   \\ 
      \cline{2-8}                                                                                                                             
      \raisebox{1.5ex}[-1.5ex]{$T_2^{+-}$}&1&    0.75(3)  &    0.60(2)  &    0.88(6)  &    0.75(5)  &    0.61(3)  &   0.533(9)   \\ 
      \hline                                                                                                                                  
      &                                    0&    1.68(13) &    1.78(8)  &     2.2(3)  &     1.9(3)  &    1.79(13) &    1.69(3)   \\ 
      \cline{2-8}                                                                                                                             
      \raisebox{1.5ex}[-1.5ex]{$T_2^{-+}$}&1&    1.87(20) &    1.59(9)  &     1.8(3)  &     1.8(4)  &    1.42(11) &    1.38(2)   \\ 
      \hline                                                                                                                                  
      &                                    0&    1.93(20) &    1.99(14) &       4(2)  &       4(2)  &    1.91(19) &    1.74(5)   \\ 
      \cline{2-8}                                                                                                                                         
      \raisebox{1.5ex}[-1.5ex]{$T_2^{--}$}&1&    0.84(6)  &    0.65(4)  &    0.68(16) &    0.63(7)  &    0.71(6)  &   0.548(13)  \\  
      \hline
    \end{tabular}
  \end{center}
\end{table}


\end{document}